\documentclass[aps,prb,onecolumn,groupedaddress,amsmath,amssymb,showpacs]{revtex4}
\usepackage{graphicx,bm,wasysym,color}

\def\be{\begin{equation}}  
\def\ee{\end{equation}}  
\def\ba{\begin{eqnarray}}  
\def\ea{\end{eqnarray}}  
\def\bc{\begin{center}}  
\def\ec{\end{center}}  
\def\p{\partial}

\begin{document}
\title{Negative dynamic conductivity of a current driven array of 
graphene nanoribbons}

\author{S. A. Mikhailov}
\email[Email: ]{sergey.mikhailov@physik.uni-augsburg.de}
\affiliation{Institute of Physics, University of Augsburg, D-86135 Augsburg, Germany}
\author{N. A. Savostianova}
\affiliation{Institute of Physics, University of Augsburg, D-86135 Augsburg, Germany}
\author{A. S. Moskalenko}
\altaffiliation{Present address: Department of Physics, University of Konstanz, Germany}
\affiliation{Institute of Physics, University of Augsburg, D-86135 Augsburg, Germany}

\date{\today}

\begin{abstract}
We consider a periodic array of graphene nanoribbons under the action of a strong dc electric field $E_0$ and an external electromagnetic excitation with the frequency $\omega$ and the lateral wave vector $q$. Solving the quasi-classical Boltzmann kinetic equation and calculating the surface dynamic conductivity $\sigma_{2D}(q,\omega,E_0)$ and the absorption coefficient of such a system we show that the real part of the conductivity and the absorption coefficient may become negative under certain conditions. Physically this corresponds to the amplification of the electromagnetic waves at the expense of the energy of the direct current source. The results are discussed in connection with experiments on the surface acoustic waves and on the Smith-Purcell-type graphene-based terahertz emitter. 
\end{abstract}


\maketitle


\section{Introduction}

The nonlinear electrodynamic and optical properties of graphene currently attract much attention and continuously growing interest. It was theoretically predicted \cite{Mikhailov07e} that the linear energy dispersion of graphene electrons should lead to a strongly nonlinear electrodynamic response of this material. This prediction was experimentally confirmed, both at microwave \cite{Dragoman10,Hotopan11} and optical \cite{Hendry10,Wu11,Gu12,Zhang12,Bykov12,Kumar13,Hong13,An14} frequencies. The nonlinear effects, such as the harmonics generation \cite{Dragoman10,Kumar13,Hong13,Bykov12,An14}, the four-wave mixing \cite{Hendry10,Hotopan11,Gu12}, the Kerr effect \cite{Wu11,Zhang12}, and others have been observed. The nonlinear parameters of graphene were found to be several orders of magnitude larger than in many nonlinear materials.

Theoretical works (e.g. Refs.  \cite{Mikhailov11c,Yao13,Cheng14a,Cheng14b,Peres14,Smirnova14,Yao14,CoxAbajo14,Cheng15,Cheng16,Sabbaghi15,Savostianova15,CoxAbajo15,Mikhailov16a,CoxAbajo16,MariniAbajo16}) predicted interesting physical phenomena, many of which have not yet been experimentally studied in details. Among them, for example, a resonant enhancement of the second harmonic due to plasma resonances in a graphene layer \cite{Mikhailov11c}, a giant nonlinear optical response of graphene in a magnetic field \cite{Yao13}, an optical bistability at terahertz frequencies \cite{Peres14}, a nonlinear generation of two-dimensional (2D) plasmons in graphene \cite{Yao14}, a direct current induced second harmonic generation at optical frequencies \cite{Cheng14b}, nonlinear plasmonic effects \cite{CoxAbajo14,CoxAbajo15,CoxAbajo16}, a resonant enhancement of the third-order nonlinear effects due to the interband optical transitions in graphene \cite{Cheng14a,Cheng15,Mikhailov16a}, a saturable absorption effect \cite{Cheng15,Mikhailov16a,MariniAbajo16} and other effects. 
These theoretical and experimental results suggest that graphene is a very promising nonlinear medium and pave the way for graphene-based nonlinear optoelectronics and photonics.  

In this paper we develop a quasi-classical theory of the low-frequency (microwave, terahertz) electrodynamic response of an array of graphene nanoribbons driven by a strong dc electric field $E_0$. Within the relaxation time approximation, we calculate the dc field dependent dynamic conductivity of a single and of an array of nanoribbons as a function of the frequency $\omega$, wave vector $q\equiv q_x$ of the external radiation ($x$ is the direction along the ribbons), of the applied dc electric field $E_0$ (which is not necessarily weak), as well as of the chemical potential and temperature. We show that the real part of the conductivity, as well as the absorption coefficient of the structure, may become negative under certain conditions, which means the amplification of waves. The results are discussed in view of their possible application for the design of a current-driven tunable graphene-based terahertz emitter\cite{Mikhailov13a,Moskalenko14}, as well as in view of the surface acoustic wave experiments in 2D crystals  \cite{Bandhu13,Bandhu14,Preciado15}.

The effect of the direct current induced negative dynamic conductivity was studied in the past, both in one-dimensional (quantum wires, e.g. Ref. \cite{Sablikov96}) and two-dimensional (GaN quantum wells, e.g. Ref. \cite{Kim04}) electron systems. In contrast to our paper, in all these works ``conventional'' electrons (with the parabolic energy dispersion) were studied.

The paper is organized as follows. In Section \ref{sec:NoDCcurrent} we consider an array of graphene nanoribbons under the action of only an inhomogeneous ac electric field (i.e., at $E_0=0$). We introduce the method of solving the problem and analyze the electromagnetic response of such an unbiased electron system. Then, in Sections \ref{sec:DCresponse} and \ref{sec:DCcurrent} we switch on the dc electric field $E_0$ and study the dc response of such a current-driven electron system (Section \ref{sec:DCresponse}) and its ac response to a weak electromagnetic excitation (Section \ref{sec:DCcurrent}). The influence of the dc electric field $E_0$ is taken into account non-perturbatively. In Section \ref{sec:conclusions} we summarize our results and draw conclusions. Some technical details of calculations are given in the Appendix \ref{app:integrals}.

\section{Electrodynamic response of a non-driven system\label{sec:NoDCcurrent}}

\subsection{Formulation of the problem\label{sec:form_prob}}

\begin{figure}
\includegraphics[width=0.49\textwidth]{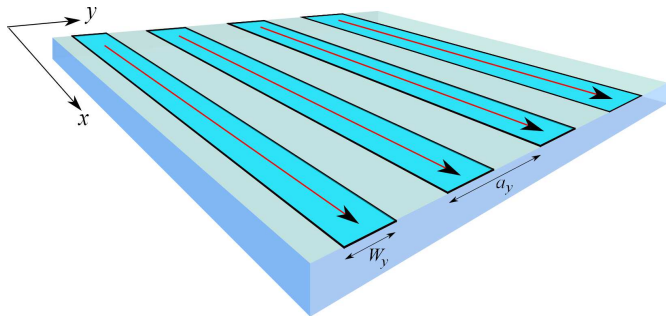}
\caption{Geometry of the graphene nanoribbon array on a substrate. Arrows show the direction of the electron motion in the ribbons.}\label{fig:geom}
\end{figure}

We assume that an array of narrow graphene nanoribbons lies on a substrate with the dielectric constant $\kappa_s$ in the plane $z=0$, Figure \ref{fig:geom}. The nanoribbons are infinite in the $x$-direction and have the width $W_y$ in the $y$-direction. The period of the structure in the $y$-direction is $a_y$. We assume that a sufficiently strong dc electric field $E_0$ is applied to the structure and a stationary charge current flows along the ribbons in the $x$-direction. In addition, a space- and time-dependent electric field $E_1(x,t)\propto \exp(iqx-i\omega t)$ influences the system, so that the total external field acting on electrons in the ribbons amounts to  
\be 
E_x(x,t)=E_0+E_1(x,t).\label{extfield}
\ee
Our goal is to calculate the linear response of the system to the ac electric field $E_1(x,t)$ \textit{not assuming} that the dc field $E_0$ is weak, i.e. the response of the system to the dc field $E_0$ is aimed to be taken into account non-perturbatively. 

The energy spectrum of electrons in two-dimensional graphene is described by the known linear dispersion relation ${\cal E}(p_x,p_y)=\pm v_F\sqrt{p_x^2+p_y^2}$ where $v_F\approx 10^8$ cm/s is the Fermi velocity in graphene\cite{Neto09}, ${\bf p}=(p_x,p_y)=\hbar(k_x,k_y)$ is the quasi-momentum, and the upper (lower) sign corresponds to the electron (hole) energy band. We assume that in a single nanoribbon with the width $W_y$ the momentum $p_y$ is quantized, so that the quasi-one-dimensional spectrum of electrons and holes has the form
\be 
{\cal E}_{k,\pm}(p_x)=\pm \sqrt{\Delta_0^2k^2+(v_Fp_x)^2},
\label{spectrum}
\ee 
where $k$ is the subband index ($k=1,2,\dots$) and 
\be
\Delta_0=\frac{\pi \hbar v_F} {2W_y}
\label{Delta0}
\ee
is the half of the band gap in the ribbon \cite{Yang07}. 

If the dc and ac electric fields (\ref{extfield}) are applied to the system of nanoribbons, its response can be described by quasi-classical Boltzmann equations written for the distribution functions of electrons and holes in all occupied subbands (we consider the low-frequency range, $\hbar\omega\lesssim E_F$, when the interband optical transitions can be neglected). Our previous studies \cite{Mikhailov98c,Mikhailov13a,Moskalenko14} showed  that, in order to realize the negative conductivity of the system, the density of electrons should be sufficiently low (this conclusion is also confirmed by the present work). Almost everywhere in this paper we will therefore assume that the chemical potential lies in the upper half of the forbidden band, $0<\mu_0\lesssim \Delta_0$, and the temperature is low, $T\ll \Delta_0$. Under these conditions one may consider only the response of electrons of the lowest ($k=1$) conduction ($+$) subband and ignore the inter-subband scattering. Then the Boltzmann equation for electrons of the $(1+)$ subband assumes the form (we omit the indexes $k,\pm$)
\be  \frac{\partial f(p_x,x,t)}{\partial t}+v_x\frac{\partial f(p_x,x,t)}{\partial x} -\Big(eE_0+eE_1(x,t)\Big)\frac{\partial f(p_x,x,t)}{\partial  p_x}=\mathrm{St}\{f\},\label{BoltEq}\ee
where $e>0$ is the elementary charge, the dc field is assumed to be negative, $E_0<0$, $v_x=\p {\cal E}/\p p_x$, and $\mathrm{St}\{f\}$ is the scattering integral. The latter describes the scattering of graphene electrons by impurities, phonons, grain boundaries, etc., and is, in general, a complicated functional of the distribution function $f(p_x,x,t)$. Since we aim to get a non-perturbative (in $|E_0|$) solution of the problem we will assume a simple relaxation time ($\tau$) model for the scattering integral,
\be 
\mathrm{St}\{f\}=-\frac{f(p_x,x,t)-f_{\mathrm{le}}(p_x,x,t)}{\tau}.
\label{scatint}
\ee
This approximation for $\mathrm{St}\{f\}$ is sufficiently reasonable and allows us to get an exact analytical solution of the response problem at a strong driving dc electric field. The function $f_{\mathrm{le}}(p_x,x,t)$ in Eq. (\ref{scatint}) is the Fermi distribution function describing the \textit{local equilibrium} \cite{Mermin70,Kragler80}
\begin{equation}
f_{\mathrm{le}}(p_x,x,t)=\frac{1}{1+\exp\left(\frac{\mathcal{E} (p_x) -\mu(x,t)}{T}\right)};
\label{LocEquil}
\end{equation}
it differs from the global equilibrium distribution
\be 
f_{\mathrm{eq}}({\cal E}(p_x))=\frac 1{1+e^{({\cal E}(p_x)-\mu_0)/T}}\label{fermi}
\ee
by the space and time dependence of the \textit{local} chemical potential $\mu(x,t)$. The local chemical potential is determined from the particle conservation condition \cite{Mermin70,Kragler80}
\be 
\sum_{p_x}\Big(f(p_x,x,t)-f_{\mathrm{le}}(p_x,x,t)\Big)=0;
\label{PartConserv}
\ee
the global chemical potential is denoted as $\mu_0$.

In the current Section we solve the Boltzmann equation (\ref{BoltEq})--(\ref{scatint})  and calculate the dynamic conductivity of the graphene nanoribbons at $E_0=0$. Then in Sections \ref{sec:DCresponse} and \ref{sec:DCcurrent} we analyze the case of a finite dc field $E_0\neq 0$.

\subsection{Parameters\label{sec:param}}

In this paper we utilize the value of $\Delta_0$, Eq. (\ref{Delta0}), as the energy scale and measure the frequency, wave vector, scattering rate, chemical potential, temperature and the electric field in the following dimensionless units
\be 
\Omega=\frac{\hbar\omega}{\Delta_0},\ \ \ 
Q=\frac{\hbar qv_F}{\Delta_0}\equiv\frac{2qW_y}\pi,\ \ \ 
\Gamma=\frac{\hbar\gamma}{\Delta_0}=\frac{\hbar}{\tau\Delta_0}=\frac{ 2}{\pi}\frac{W_y}{l}, \ \ \ \tilde \mu=\frac {\mu_0}{\Delta_0}, \ \ \ \tilde T=\frac T{\Delta_0},\ \ \ {\cal F}=\frac{e(-E_0)v_F\tau}{\Delta_0}=\frac{e|E_0|l}{\Delta_0},
\label{param}
\ee
where $l=v_F\tau$ is the mean free path and $\gamma=1/\tau$ is the scattering rate. Consider an illustrative numerical example. Assume that the ribbon width is $W_y=400$ nm. Then the value of the energy scale is $\Delta_0\approx 2.45$ meV. This is equivalent to $\approx 28$ K and $\approx 0.62$ THz. The temperature parameter $\tilde T\simeq 0.2$ thus corresponds to $T\simeq 5$ K, and the dimensionless frequency $\Omega\approx 2$ to about $1.2$ THz. 
The spatial periodicity of the external ac electric field can be created by placing a grating (with the period $a_x$ in the $x$-direction) in the vicinity of the nanoribbon array. Then $q\simeq 2\pi/a_x$ and the dimensionless wave-vector parameter $Q\simeq 4W_y/a_x$ can vary from values small as compared to unity up to $Q\simeq 4$ and above. The value of $Q\simeq 2$ corresponds to the grating period $a_x\simeq 800$ nm. 

The scattering parameter $\Gamma$ depends on the sample quality. With $W=400$ nm and the mean free path about 1.3 $\mu$m, the factor $\Gamma$ constitutes about $0.2$. 
The field parameter ${\cal F}$ reaches 1 at $|E_0|\simeq 19$ V/cm in the samples with the indicated value of the mean free path. At $|E_0|\simeq 100$ V/cm $=10$ mV/$\mu$m we have ${\cal F}\simeq 5$.

\subsection{Equilibrium density\label{sec:eqdens}}

\begin{figure}
\includegraphics[width=0.6\textwidth]{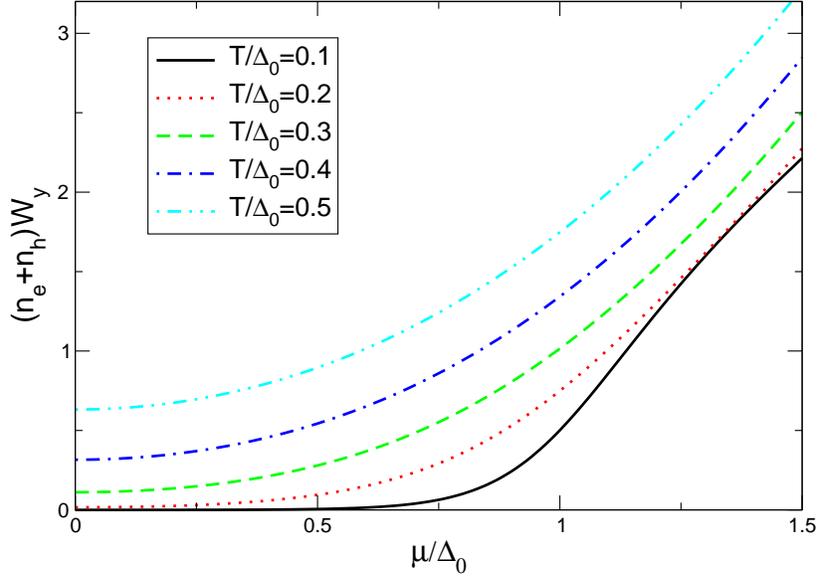}
\caption{The 1D density of charge carriers (electrons and holes) as a function of the chemical potential $\tilde\mu=\mu/\Delta_0$ at different temperatures $T/\Delta_0$. }\label{density}
\end{figure}

Now we consider the equilibrium linear density of charge carriers (electrons $n_e$ and holes $n_h$) as a function of the chemical potential at different temperatures. The result reads (here, in Section \ref{sec:eqdens} only, we take into account all electron and hole subbands)
\be 
n_e+n_h=
\frac{g_sg_v} {2W_y}\sum_{k=1}^\infty\Big({\cal N}_k(\tilde\mu,\tilde T)+{\cal N}_k(-\tilde\mu,\tilde T)\Big),
\ee
where $g_s=g_v=2$ are the spin and valley degeneracies, and 
\be 
{\cal N}_k(\tilde\mu,\tilde T)=
\int_{0}^\infty 
\frac{dP}{1+\exp\left(\frac{\sqrt{k^2+P^2}-\tilde \mu}
{\tilde T}\right)}=k\int_{0}^1 \frac {dV}{(1-V^2)^{3/2}} \frac { 1}{1+\exp \frac{k/\sqrt{1-V^2}-\tilde\mu }{\tilde T }  }.
\ee
The $\tilde \mu$-dependence of the total dimensionless charge density $(n_e+n_h)W_y$ is shown in Figure \ref{density}. At $\tilde T\simeq 0.2$ and $\tilde \mu\approx 1$ the parameter $n_lW_y\equiv(n_e+n_h)W_y\simeq 0.75$; under these conditions the main contribution to the density is given by electrons of the lowest subband. If $W_y=400$ nm, the linear charge density $n_l$ is about $1.875\times 10^4$/cm. If the aspect ratio $W_y/a_y$ of the nanoribbon array is about $1/2$ the two-dimensional (2D) average charge carrier density $n_s$ is 
\be 
n_s=\frac{n_l}{a_y}\approx 2.34\times 10^{8}\ \rm{cm}^{-2}.\label{dens-estimate}
\ee
It will be seen below that the amplification of the waves in the current driven system of graphene nanoribbons can be achieved at low electron densities. The number (\ref{dens-estimate}) gives a typical scale of the charge carrier density needed for the realization of the amplification.

\subsection{Solution of the Boltzmann equation}

In the absence of the dc electric field $E_0$ the Boltzmann equation (\ref{BoltEq})--(\ref{scatint}) reads

\begin{equation}\label{Eq:Boltzmann_le0}
  \frac{\partial f(p_x,x,t)}{\partial t}+
v_x\frac{\partial f(p_x,x,t)}{\partial x}
-e E_1(x,t)\frac{\partial f(p_x,x,t)}{\partial
  p_x}=-\frac{f(p_x,x,t)-f_{\mathrm{le}}(p_x,x,t)}{\tau}.
\end{equation}
Substituting 
\begin{equation}\label{Eq:f_Antsatz}
   f(p_x,x,t)=f_{\mathrm{eq}}({\cal E})+ f_1(p_x,x,t)
\end{equation}
and
\begin{equation}\label{Eq:f_eq_le_approx}
   f_{\mathrm{le}}(p_x,x,t)= f_{\mathrm{eq}}({\cal E})-\frac{\partial f_{\mathrm{eq}}({\cal E})}{\partial \mathcal{E}}\mu_1(x,t)
\end{equation}
into Eq. (\ref{Eq:Boltzmann_le0}), where $\mu_1(x,t)=\mu(x,t)-\mu_0$, we get the linearized (in $E_1$) version of the Boltzmann equation 
\begin{equation}\label{Eq:Boltzmann_le0b}
  \frac{\partial f_1(p_x,x,t)}{\partial t}+v_x\frac{\partial f_1(p_x,x,t)}{\partial x}
-e E_1(x,t)\frac{\partial f_{\mathrm{eq}}({\cal E})}{\partial
  p_x}=-\gamma\left(f_1(p_x,x,t)+\frac{\partial f_{\mathrm{eq}}({\cal E})}{\partial \mathcal{E}}\mu_1(x,t)\right).
\end{equation}
From  the charge-density conservation condition (\ref{PartConserv}) we get
\be 
\mu_1(x,t)=-\frac{\sum_{p_x'}f_1(p_x',x,t)}{\sum_{p_x'}\frac{\partial f_{\mathrm{eq}}(\mathcal{E}')}
{\partial \mathcal{E}'}}.
\ee

The solution of the linearized Boltzmann equation (\ref{Eq:Boltzmann_le0b}) is searched for in the form $\propto e^{iqx-i\omega t}$. Then we obtain
\begin{equation}
 f_1(p_x)
-\frac{i\gamma}{\omega  +i\gamma -q v_x  }\frac{\frac{\partial f_{\mathrm{eq}}(\mathcal{E})}
{\partial \mathcal{E}}}
{\sum_{p_x'}\frac{\partial f_{\mathrm{eq}}(\mathcal{E}')}
{\partial \mathcal{E}'}}\sum_{p_x'}f_1(p_x')
=i\frac{e E_1v_x}{\omega  +i\gamma -q v_x  }
\frac{\partial f_{\mathrm{eq}}(\mathcal{E})}{\partial
 \mathcal{E} }.\label{eq:1}
\end{equation}
The second term in the left-hand side contains an unknown constant $\sum_{p_x'}f_1(p_x')$. To find it we perform summation over $p_x$ in both sides of Eq. (\ref{eq:1}) and get
\begin{equation}
\sum_{p_x} f_1(p_x)
=ie E_1
\left(\sum_{p_x}\frac{\partial f_{\mathrm{eq}}(\mathcal{E})}
{\partial \mathcal{E}}\right)
\frac{\left(
\sum_{p_x}\frac{v_x}{\omega  +i\gamma -q v_x  }
\frac{\partial f_{\mathrm{eq}}(\mathcal{E})}{\partial
 \mathcal{E} }\right)}
{\left(
 \sum_{p_x}\frac{\omega  -q v_x}{\omega  +i\gamma -q v_x  }
\frac{\partial f_{\mathrm{eq}}(\mathcal{E})}
{\partial \mathcal{E}}\right)}.
\end{equation}
The solution of the Boltzmann equation then assumes the form
\be
 f_1(p_x)=
\frac{ie E_1}{\omega  +i\gamma -q v_x}\frac{\partial f_{\mathrm{eq}}(\mathcal{E})}
{\partial \mathcal{E}}
\left(v_x
 + \frac{i\gamma
\sum_{p_x'}\frac{v_x'}{\omega  +i\gamma -q v_x'  }
\frac{\partial f_{\mathrm{eq}}(\mathcal{E}')}{\partial
 \mathcal{E}' }}
{\sum_{p_x'}
\frac{\omega   -q v_x' }{\omega  +i\gamma -q v_x' } \frac{\partial f_{\mathrm{eq}}(\mathcal{E}')}
{\partial \mathcal{E}'}
}
\right).
\ee

\subsection{Current density }

Calculating the current according to the standard formula  
\be 
j_1=-e\frac{g_sg_v}{L}\sum_{p_x}v_xf_1(p_x),
\label{1d_current}
\ee
where $L$ is the ribbon length in the $x$-direction, we obtain
\be 
j_1=
-ie^2 E_1
\left({\cal A}_2(\omega,\gamma,q,\mu,T)
 + i\gamma
\frac{{\cal A}^2_1(\omega,\gamma,q,\mu,T)
}
{\omega {\cal A}_0(\omega,\gamma,q,\mu,T)-
q{\cal A}_1(\omega,\gamma,q,\mu,T) }
\right).
\ee
Here we have defined the integrals 
\be 
{\cal A}_n(\omega,\gamma,q,\mu,T)=
\frac{g_sg_v}{L}\sum_{p_x}
\frac{\partial f_{\mathrm{eq}}(\mathcal{E})}
{\partial \mathcal{E}}
\frac{v_x^n }{\omega  +i\gamma -q v_x }.
\ee
After some transformations they can be presented in the following form
\be
{\cal A}_n(\omega,\gamma,q,\mu,T)=
i\frac{g_sg_v}{2\pi}\frac{v_F^{n-1} }{2\Delta_0}
 {\cal I}_n(\Omega,Q,\Gamma,\tilde \mu,\tilde T),
\ee
with the dimensionless integrals 
\be
{\cal I}_n(\Omega,Q,\Gamma,\tilde \mu,\tilde T)=
\frac i{2\tilde T}
\int_{-1}^1 \frac{dV}
{\Omega  +i\Gamma -QV }\frac{V^{n}}{(1-V^2)^{3/2}}
\cosh^{-2}\left(\frac{\frac 1{\sqrt{1-V^2}}-\tilde\mu}{2\tilde T}\right).
\ee
${\cal I}_n(\Omega,Q,\Gamma,\tilde \mu,\tilde T)$ can be evaluated numerically as functions of all their dimensionless arguments.

\subsection{Dynamic conductivity}

The current $j_1$ in Eq. (\ref{1d_current}) is a one-dimensional (1D) current flowing in a single nanoribbon. The corresponding 1D dynamic conductivity is $\sigma_{1D}(q,\omega)=j_1/E_1$. In an array of nanoribbons it is reasonable to determine the 2D conductivity as an average current density flowing in all nanoribbons divided by the electric field. It is related to the 1D conductivity as  
\be 
\sigma_{2D}(q,\omega)=\sigma_{1D}(q,\omega)/a_y.
\label{2d-conduct-def}
\ee
Then we get for $ \sigma_{2D}(q,\omega)$:
\be 
\sigma_{2D}(q,\omega)=\frac{e^2}{\pi\hbar}\frac{g_sg_v}{2\pi } 
\frac { W_y}{  a_y}  \,
 {\cal S}_0(\Omega,Q,\Gamma,\tilde \mu,\tilde T),
\label{2d-conduct}
\ee
where
\be 
 {\cal S}_0(\Omega,Q,\Gamma,\tilde \mu,\tilde T)=
 {\cal I}_2(\Omega,Q,\Gamma,\tilde \mu,\tilde T)
 + 
\frac{i \Gamma
 {\cal I}_1^2(\Omega,Q,\Gamma,\tilde \mu,\tilde T) }
{\Omega 
 {\cal I}_0(\Omega,Q,\Gamma,\tilde \mu,\tilde T)- Q
 {\cal I}_1(\Omega,Q,\Gamma,\tilde \mu,\tilde T) } .
\label{calSfunc}
\ee
If, instead of one, the system consists of several ($N$) parallel graphene nanoribbon layers, the right hand side of Eq. (\ref{2d-conduct}) should be multiplied by $N$. 

\begin{figure}
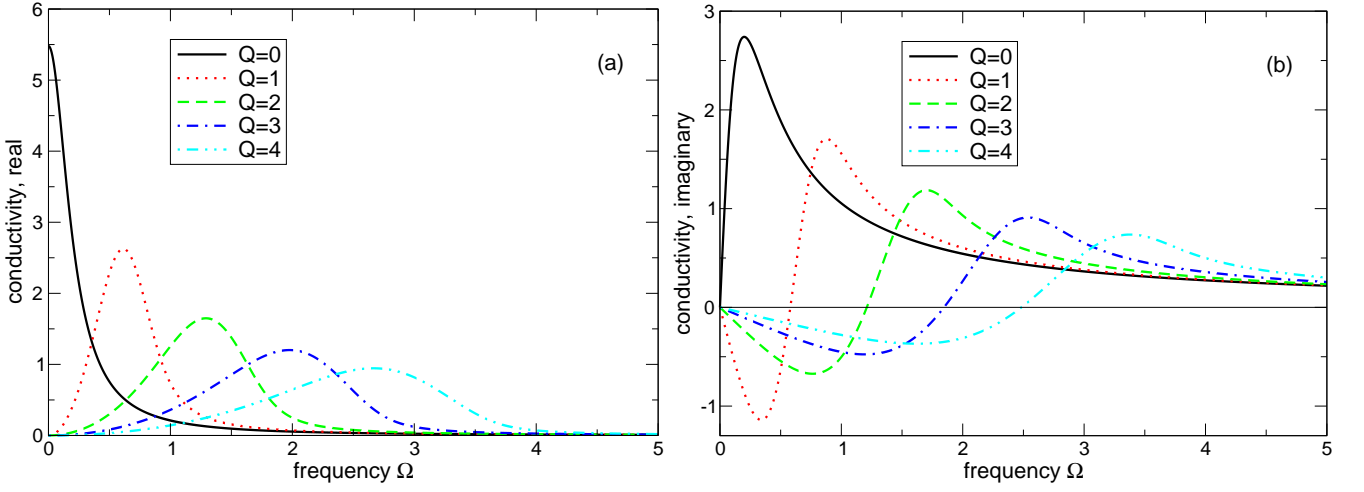

\includegraphics[width=0.49\textwidth]{fig3a.eps}
\includegraphics[width=0.49\textwidth]{fig3b.eps}
\caption{The (a) real and (b) imaginary parts of the function ${\cal S}_0(\Omega,Q,\Gamma,\tilde\mu,\tilde T)$, Eq. (\ref{2d-conduct}), vs frequency $\Omega$ at $\mu/\Delta_0=1$, $\Gamma=0.2$, temperature $T/\Delta_0=0.2$ and several values of the wave vector $Q$. The dc electric field is zero, the number of 2D layers $N=1$.
\label{fig:conductivity}}
\end{figure}

Figure \ref{fig:conductivity} shows the frequency dependence of the dimensionless conductivity $ {\cal S}_0(\Omega,Q,\Gamma,\tilde \mu,\tilde T)$ from Eq. (\ref{2d-conduct}) at different wave vectors. The real (Fig. \ref{fig:conductivity}(a)) and imaginary (Fig. \ref{fig:conductivity}(b)) parts of the conductivity are even and odd functions of $\Omega$, respectively.  At $Q=0$ Eq. (\ref{2d-conduct}) is reduced to the Drude formula (black solid curves in Figure \ref{fig:conductivity}). At a larger $Q$ the maximum is shifted to larger values of $\Omega$ and becomes broader. The maxima are located at $\Omega<Q$ due to the single particle absorption caused by electrons moving with the velocity smaller than $v_F$. The imaginary part of the conductivity, Figure \ref{fig:conductivity}(b), is connected to the real part by the Kramers-Kronig relations.

\subsection{Dielectric function\label{noDCdielfunc}}

\begin{figure}
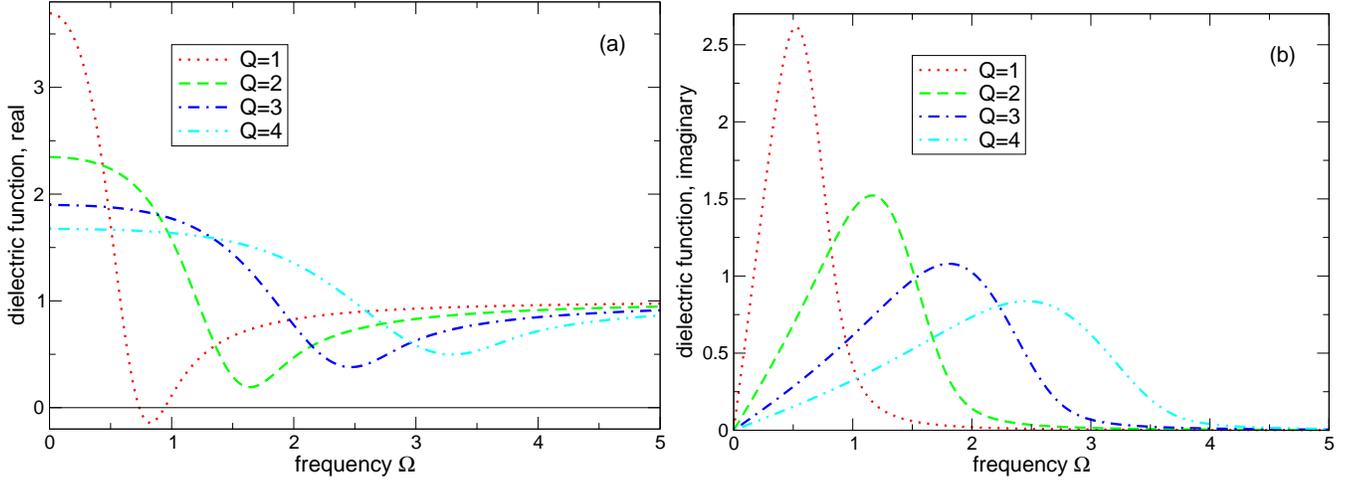

\includegraphics[width=0.49\textwidth]{fig4a.eps}
\includegraphics[width=0.49\textwidth]{fig4b.eps}
\caption{The (a) real and (b) imaginary parts of the dielectric function  $\epsilon_{2D}(q,\omega)$ vs frequency at $ \mu/\Delta_0=1$, $\Gamma=0.2$,  temperature $ T/\Delta_0=0.2$ and several values of the wave vector $Q$. The dc electric field is zero, the number of 2D layers $N=1$, $\kappa=2.45$, $W_y/a_y=0.5$.
\label{fig:dielfunc}}
\end{figure}

If the 2D layer with the conductivity $\sigma_{2D}(q,\omega)$ lies on the dielectric substrate with the dielectric constant $\kappa_s$, the effective  dielectric function of such a system can be determined as \cite{Stern67} 
\be 
\epsilon_{2D}(q,\omega)=1+\frac{2\pi i \sigma_{2D}(q,\omega)} {\omega \kappa}|q|,
\ee 
where $\kappa=(\kappa_s+1)/2$ is the effective dielectric constant of the surrounding medium. Substituting the expression for the conductivity (\ref{2d-conduct}) we get 
\be 
\epsilon_{2D}(q,\omega)=
1+i\frac{ g_sg_v }{\pi  }\frac{|Q|}{\Omega}\frac{e^2}{\hbar v_F\kappa}\frac {  W_y}{  a_y}{\cal S}_0(\Omega,Q,\Gamma,\tilde \mu,\tilde T).\label{dielfunc}
\ee
The real and imaginary parts of the dielectric function (\ref{dielfunc}) are shown in Figure \ref{fig:dielfunc}.   

\subsection{Absorption\label{noDCabsorption}}

If the potential of the external electric field is $\phi^{ext}_{q\omega}$, the potential of the total electric field acting on graphene electrons is $\phi^{tot}_{q\omega}=\phi^{ext}_{q\omega}/\epsilon(q,\omega)$, and the induced linear-response current reads 
\be 
j_{q\omega}=\frac{\sigma_{2D}(q,\omega)} {\epsilon(q,\omega)}E^{ext}_{q\omega}.
\ee
The time-averaged Joule heat is then given by $\textrm{Re\,}\sigma_{2D}(q,\omega)|E^{ext}_{q\omega}|^2/2|\epsilon(q,\omega)|^2$. If we assume that the time-averaged magnitude of the Poynting vector of the incident wave is $c|E^{ext}_{q\omega}|^2/8\pi$, the absorption coefficient can be found as 
\be 
{\cal A}=\frac{4\pi}{c}\frac{\textrm{Re }\sigma_{2D}(q,\omega)} {|\epsilon_{2D}(q,\omega)|^2} ,\label{absorption}
\ee
where $\sigma_{2D}(q,\omega)$ and $\epsilon_{2D}(q,\omega)$ are given by Eqs. (\ref{2d-conduct}) and (\ref{dielfunc}). 

\begin{figure}
\includegraphics[width=0.49\textwidth]{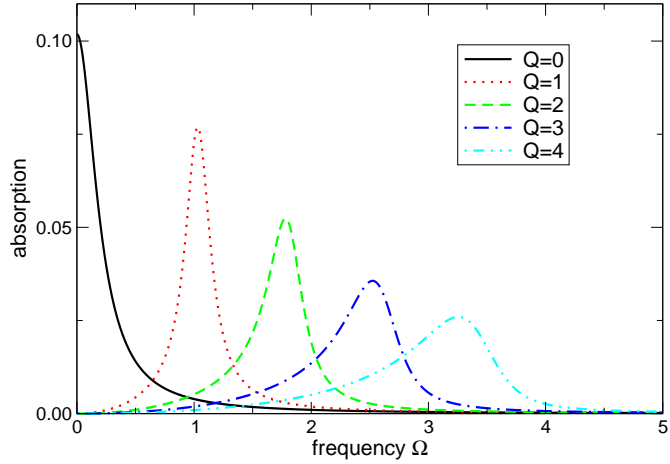}
\caption{The absorption coefficient (\ref{absorption}) vs frequency at $\mu/\Delta_0=1$, $\Gamma=0.2$,  temperature $T/\Delta_0=0.2$ and different values of $Q$. The dc electric field is zero, the number of 2D layers $N=1$.
\label{fig:absorp}}
\end{figure}

The absorption spectrum of an array of graphene nanoribbons is shown in Figure \ref{fig:absorp}. The maxima of the absorption coefficient ${\cal A}$ are shifted with respect to those of Re\,${\cal S}_0$ to higher frequencies, and the linewidths are a bit narrower. Notice that, under the chosen conditions ($\tilde\mu=1$, $\tilde T\simeq 0.2$) and at $Q\gtrsim 1$ the real part of the dielectric function $\epsilon_{2D}(q,\omega)$, Figure \ref{fig:dielfunc}(a), does not vanish at any frequency, i.e. the 2D plasmons whose spectrum satisfies the equation Re\,$[\epsilon_{2D}(q,\omega)]=0$, do not exist. The absorption maxima in Figure \ref{fig:absorp} at $Q\gtrsim 1$ thus have the single-particle origin. At smaller $Q$ these maxima correspond to the collective (2D plasmon) resonance.

\section{DC response of a current driven system\label{sec:DCresponse}}

Now we consider the case when a strong dc electric field $E_0$ is applied to the system but the amplitude of the ac field is zero, $E_1=0$. We calculate the stationary non-equilibrium distribution function which is formed under the action of the dc field $E_0$ whereas the scattering processes are described within the $\tau$-approximation (\ref{scatint}).

\subsection{Stationary distribution function}

\begin{figure}
\includegraphics[width=0.49\textwidth]{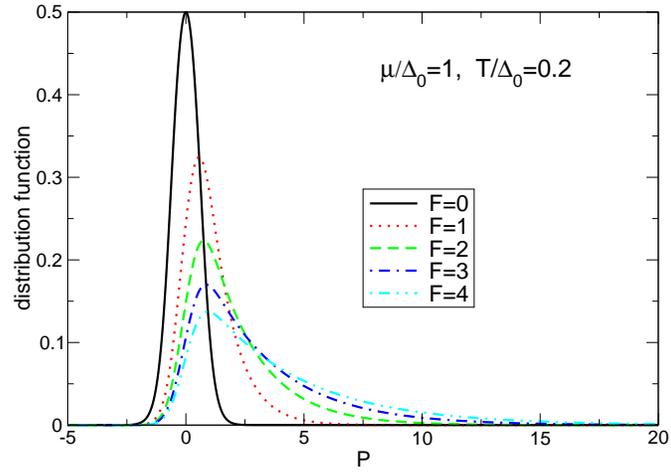}
\caption{The distribution function (\ref{f-sol}) at $\mu/\Delta_0=1$ and $T/\Delta_0=0.2$ for different values of the electric field parameter ${\cal F}$.}\label{fig:dfmu1T0.2}
\end{figure}

Under the action of the uniform and time-independent electric field $E_0$ the Boltzmann equation for the stationary distribution function $f_0(p_x)$ reads (we assume $e>0$ and $E_0<0$):
\be 
\frac{d f_0(p_x)}{d p_x}=\frac{f_0(p_x)-f_{\mathrm{eq}}(p_x)}{eE_0\tau} .
\label{statcase}
\ee
It is solved by the separation of variables method; the solution can be written in the following dimensionless form
\be 
f_0(P,\tilde\mu,\tilde T,{\cal F})=
\int_{0}^\infty  \frac{e^{-x}} {1+\exp\left(\frac{\sqrt{1+(P-{\cal F} x)^2}-\tilde\mu } {\tilde T }\right)}dx ,
\label{f-sol}
\ee
where we have used the boundary condition $f_0(p_x)\to 0$ at $p_x\to\pm\infty$ and introduced the normalized momentum $P=v_Fp_x/\Delta_0$. The function (\ref{f-sol}) satisfies the condition
\be 
f_0(P,\tilde\mu,\tilde T,-{\cal F})=f_0(-P,\tilde\mu,\tilde T,{\cal F})
\ee 
and is shown in Figure \ref{fig:dfmu1T0.2}. At $e>0$ and $E_0<0$ electrons move in the positive $x$-direction. One sees that the stationary distribution function is strongly asymmetric and substantially differs from the equilibrium distribution function already at ${\cal F}\gtrsim 1$. 

\subsection{Stationary dc current and average drift velocity}

\begin{figure}
\includegraphics[width=0.49\textwidth]{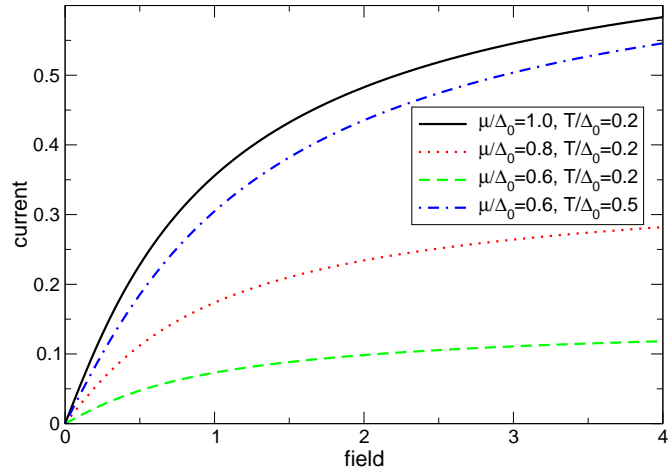}
\caption{The current (\ref{dccurrent}) vs the dc electric field ${\cal F}$ at different values of the chemical potential $\tilde\mu=\mu/\Delta_0$ and the temperature $\tilde T=T/\Delta_0$ .}\label{fig:dccurrent}
\end{figure}

Having obtained the stationary distribution function we now calculate the dc current density (\ref{1d_current}). It can be reduced to the dimensionless form
\be
j_0=
-\frac{g_sg_v}{4 }\frac{e v_F} {W_y}{\cal J}(\tilde\mu,\tilde T,{\cal F}) ,
\ee
where 
\be 
{\cal J}(\tilde\mu,\tilde T,{\cal F})=
\int_{-\infty}^\infty \frac{PdP}{\sqrt{1+P^2}}  f_0(P,\tilde\mu,\tilde T,{\cal F}) =
\int_{-1}^1 \frac{VdV}{(1-V^2)^{3/2}}  
f_0\left(\frac V{\sqrt{1-V^2}},\tilde\mu,\tilde T,{\cal F}\right).
\label{dccurrent}
\ee
The dc-field dependence of the current for several values of the chemical potential and temperature is illustrated in Figure \ref{fig:dccurrent}. The current-voltage characteristics is linear only at ${\cal F}\ll 1$. At larger values of ${\cal F}$ one sees substantial deviations from the Ohm's law. 

\begin{figure}
\includegraphics[width=0.49\textwidth]{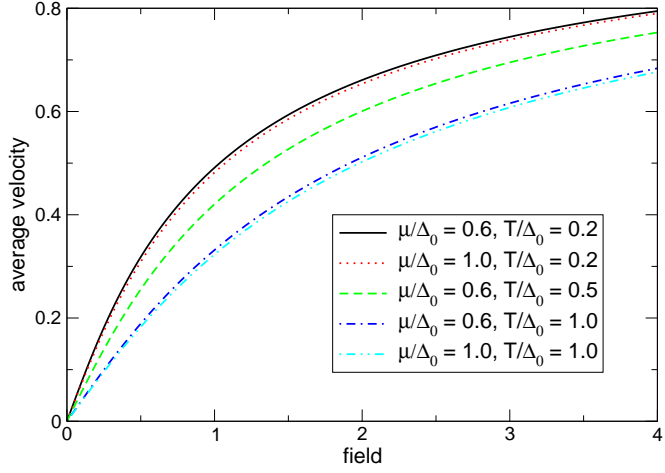}
\caption{The average drift velocity of electrons (\ref{aveveloc}) vs the dc electric field ${\cal F}$ at different values of the chemical potential $\tilde\mu=\mu/\Delta_0$ and the temperature $\tilde T=T/\Delta_0$. }\label{fig:aveveloc}
\end{figure}

The average drift velocity of electrons in the first electron subband, 
\be 
V_{\rm dr}(\tilde\mu,\tilde T,{\cal F})\equiv \frac{\bar v_{\rm dr}(\tilde\mu,\tilde T,{\cal F})}{v_F}=\frac{\int_{-\infty}^\infty V(P)  f_0(P,\tilde\mu,\tilde T,{\cal F})dP}{\int_{-\infty}^\infty   f_0(P,\tilde\mu,\tilde T,{\cal F})dP}
=\frac{{\cal J}(\tilde\mu,\tilde T,{\cal F})}{2{\cal N}_1(\tilde\mu ,\tilde T)},
\label{aveveloc}
\ee
is shown in Figure \ref{fig:aveveloc}; here 
\be 
V(P)=\frac{P}{\sqrt{1+P^2}} \label{dimlessvelocity}
\ee
is the dimensionless velocity of electrons in the first electron subband. Notice that $\bar v_{\rm dr}(\tilde\mu,\tilde T,{\cal F})$ weakly depends on the chemical potential but is sensitive to a change of the temperature.

The $I$-$V$ characteristics similar to the one shown in Figure \ref{fig:dccurrent} have been experimentally observed in graphene, see, e.g., Fig. 2b in Ref. \cite{Barreiro09}. The average drift velocity which was achieved in that paper (at much higher electron density and temperature than we assume here) was about $0.36 v_F$ which confirms the feasibility of our model.

\section{AC response of a current driven system\label{sec:DCcurrent}}

\subsection{Dynamic distribution function in a driven system}

Now we consider the system response to the ac electric field $E_1(x,t)=E_1e^{iqx-i\omega t}$ in the presence of the strong dc driving field $E_0$. It is described by the Boltzmann equation (\ref{BoltEq})--(\ref{scatint}). We search for a solution in the form
\begin{equation}
   f(p_x,x,t)=f_{0}(p_x)+ f_1(p_x,x,t),
\end{equation}
where $f_{0}(p_x)$ is the stationary distribution function found in Section \ref{sec:DCresponse}, Eq. (\ref{statcase}), and $f_1(p_x,x,t)$ is the correction proportional to the perturbation $E_1(x,t)$. Linearizing the Boltzmann equation we get the following equation for $f_1(p_x,x,t)$:
\be
 \left( \frac{\partial }{\partial t}+
v_x\frac{\partial }{\partial x}
- eE_0\frac{\partial }{\partial  p_x}\right)f_1(p_x,x,t)
+\gamma\left(f_1(p_x,x,t)-
\frac{\frac{\partial f_{\mathrm{eq}}(\mathcal{E})}
{\partial \mathcal{E}}}
{\sum_{p_x'}\frac{\partial f_{\mathrm{eq}}(\mathcal{E}')}
{\partial \mathcal{E}'}}\sum_{p_x'}f_1(p_x',x,t)
\right)
=eE_1(x,t)\frac{\partial f_0(p_x)}{\partial  p_x}.
\ee
Since the perturbation is proportional to $e^{iqx-i\omega t}$, we search for a solution with the same spatio-temporal form $f_1(p_x,x,t)=f_1(p_x)e^{iqx-i\omega t}$ and obtain the integro-differential equation for $f_1(p_x)$:
\be
\frac{d f_1(p_x)}{d  p_x}+
\frac{i(\omega +i\gamma -qv_x)}{eE_0}  f_1(p_x)
=-\frac{\gamma}{eE_0}
\frac{\frac{\partial f_{\mathrm{eq}}(\mathcal{E}(p_x))}
{\partial \mathcal{E}}}
{\sum_{p_x'}\frac{\partial f_{\mathrm{eq}}(\mathcal{E}(p_x'))}
{\partial \mathcal{E}'}}\sum_{p_x'}f_1(p_x')
-\frac{E_1}{E_0}\frac{d f_0(p_x)}{d  p_x}.
\ee
Its general solution reads:
\be
 f_1(p_x)
=\frac{E_1}{|E_0|}
\int^{p_x}_{-\infty} \left(
\frac{d f_0(p_x')}{d  p_x'}+{\cal C}\frac{\partial f_{\mathrm{eq}}(\mathcal{E}(p_x'))}
{\partial \mathcal{E}'}\right)
\exp\left(
-\frac{1-i\omega\tau  }{e|E_0|\tau} (p_x-p_x')
-\frac{iq}{e|E_0|} \Big({\cal E}(p_x)-{\cal E}(p_x')\Big)
\right)
dp_x',
\label{df-driven}
\ee
where 
\be 
{\cal C}=\frac{\gamma}{eE_1}\frac{1}
{\sum_{p_x}\frac{\partial f_{\mathrm{eq}}(\mathcal{E}(p_x))}
{\partial \mathcal{E}}}\sum_{p_x}f_1(p_x)
\ee
is, again, an unknown constant proportional to $\sum_{p_x}f_1(p_x)$, cf. Eq. (\ref{eq:1}). It can be found by taking a sum over $p_x$ in Eq. (\ref{df-driven}):
\be 
{\cal C}
=
\frac
{\frac{\gamma}{e|E_0|}\int_{-\infty}^\infty dp_x 
\int^{p_x}_{-\infty} 
\frac{d f_0(p_x')}{d  p_x'}
\exp\left(
-\frac{1-i\omega\tau  }{e|E_0|\tau} (p_x-p_x')
-\frac{iq}{e|E_0|} \Big({\cal E}(p_x)-{\cal E}(p_x')\Big)
\right)
dp_x'}
{\int_{-\infty}^\infty dp_x\frac{\partial f_{\mathrm{eq}}(\mathcal{E}(p_x))}
{\partial \mathcal{E}}-\frac{\gamma}{e|E_0|}\int_{-\infty}^\infty dp_x 
\int^{p_x}_{-\infty} \frac{\partial f_{\mathrm{eq}}(\mathcal{E}(p_x'))}
{\partial \mathcal{E}'}
\exp\left(
-\frac{1-i\omega\tau  }{e|E_0|\tau} (p_x-p_x')
-\frac{iq}{e|E_0|} \Big({\cal E}(p_x)-{\cal E}(p_x')\Big)
\right)
dp_x'}.
\label{constC}
\ee
The formulas (\ref{df-driven}) and (\ref{constC}) give the closed-form analytical expression for the electron distribution function in a direct current driven system of graphene nanoribbons.

\subsection{Current density, field-dependent conductivity and absorption coefficient: Analytic formulas}

In order to calculate the current density $j_1$ in a dc-driven graphene nanoribbon we substitute the distribution function (\ref{df-driven}), (\ref{constC}) into the definition (\ref{1d_current}). The dc-field dependent 1D conductivity of a single nanoribbon is then given by the relation $\sigma_{1D}(q,\omega,E_0)=j_1/E_1$, and the effective 2D conductivity of an array of nanoribbons is determined by Eq. (\ref{2d-conduct-def}). Introducing dimensionless variables $P=v_Fp_x/\Delta_0$, $P'=v_Fp'_x/\Delta_0$, and (\ref{param}) we get the following expression for the 2D conductivity:
\be 
\sigma_{2D}=
\frac{e^2}{\pi\hbar}\frac{g_sg_v}{2\pi }
\frac{W_y}{  a_y}N
{\cal S}(\Omega,Q,\Gamma,\tilde \mu,\tilde T,{\cal F})
,\label{sigma2D}
\ee
where the formula is written for $N$ parallel graphene nanoribbon layers and the dimensionless function ${\cal S}(\Omega,Q,\Gamma,\tilde \mu,\tilde T,{\cal F})$ is determined by the formula
\be 
{\cal S}(\Omega,Q,\Gamma,\tilde \mu,\tilde T,{\cal F})
=
\frac{2}{{\cal F}\Gamma}\left[{\cal K}_1(\Omega,Q,\Gamma,\tilde \mu,\tilde T,{\cal F})-\frac
{{\cal K}_0(\Omega,Q,\Gamma,\tilde \mu,\tilde T,{\cal F})\,{\cal L}_1(\Omega,Q,\Gamma,\tilde \mu,\tilde T,{\cal F})}
{{\cal F}{\cal M}(\tilde \mu,\tilde T)
+{\cal L}_0(\Omega,Q,\Gamma,\tilde \mu,\tilde T,{\cal F})}\right].
\label{locSloc}
\ee
Here we have defined the following functions:
\be 
{\cal K}_n(\Omega,Q,\Gamma,\tilde \mu,\tilde T,{\cal F})=-
\int_{-\infty}^\infty V^n(P)dP
\int_{-\infty}^P
\frac{df_0(P',\tilde\mu,\tilde T,{\cal F})}{dP'}
{\cal D}(P,P';\Omega,Q,\Gamma,{\cal F})
dP' ,\label{funcKn}
\ee
\be 
{\cal L}_n(\Omega,Q,\Gamma,\tilde \mu,\tilde T,{\cal F})=-
\int_{-\infty}^\infty V^n(P)dP 
\int_{-\infty}^P
\frac {1}{\cosh^2\frac{\sqrt{1+P'^2}-\tilde\mu}{2\tilde T}}
{\cal D}(P,P';\Omega,Q,\Gamma,{\cal F})dP',\label{funcLn}
\ee
\be 
{\cal M}(\tilde \mu,\tilde T)=\int_{-\infty}^\infty 
\frac {dP}{\cosh^2\frac{\sqrt{1+P^2}-\tilde\mu}{2\tilde T}};\label{funcM}
\ee
where 
\be 
{\cal D}(P,P';\Omega,Q,\Gamma,{\cal F})=
\exp\left(i\frac{(\Omega+i\Gamma)(P-P')-Q\left(\sqrt{1+P^2}-\sqrt{1+P'^2}\right)}{{\cal F}\Gamma}
\right)\label{funcD-exponent}.
\ee
Equations (\ref{locSloc}) -- (\ref{funcD-exponent}) determine the current-driven nonlocal frequency-dependent conductivity of an array of graphene nanoribbons as a function of the frequency $\Omega$, wave-vector $Q$, scattering rate $\Gamma$, chemical potential $\tilde \mu$, temperature $\tilde T$ and the driving dc electric field ${\cal F}$. Without the driving field, ${\cal F}=0$, the function ${\cal S}$ is reduced to ${\cal S}_0$,
\be 
{\cal S}(\Omega,Q,\Gamma,\tilde \mu,\tilde T,{\cal F}=0)
={\cal S}_0(\Omega,Q,\Gamma,\tilde \mu,\tilde T)
\ee
defined in Eq. (\ref{calSfunc}).

The effective dielectric function is introduced like in Section \ref{noDCdielfunc}; it is now related to the function ${\cal S}$:
\be 
\epsilon_{2D}(q,\omega)=
1+i\frac{ g_sg_v }{\pi  }\frac{|Q|}{\Omega}\frac{e^2}{\hbar v_F\kappa}\frac {  W_y}{  a_y} N {\cal S}(\Omega,Q,\Gamma,\tilde \mu,\tilde T,{\cal F}).\label{dielfuncDriven}
\ee
The formula for the absorption coefficient ${\cal A}$ in the current driven system of graphene nanoribbons is given by Eq. (\ref{absorption}), in which the function $\sigma_{2D}(q,\omega)$ and $\epsilon_{2D}(q,\omega)$ should be taken from Eqs. (\ref{sigma2D}) and (\ref{dielfuncDriven}) respectively. 

The formulas (\ref{sigma2D}), (\ref{dielfuncDriven}) and (\ref{absorption}) determine the electrodynamic response of the system of graphene nanoribbons driven by a strong dc current. In order to proceed further and plot the corresponding figures we have to numerically calculate the integrals ${\cal K}_n$, ${\cal L}_n$ and ${\cal M}$ in Eqs. (\ref{funcKn}) -- (\ref{funcM}). The integral (\ref{funcM}) can be easily calculated since the integrand of this function is a localized function of $P$. The integrals ${\cal L}_n$ and, especially, ${\cal K}_n$, require a careful treatment since they are, in fact, a double (${\cal L}_n$) and triple (${\cal K}_n$) integrals with strongly oscillating (due to the exponential function ${\cal D}$) integrands. The method utilized for numerical evaluation of such integrals is briefly described in Appendix \ref{app:integrals}. 

\subsection{Dynamic conductivity and absorption: Results}

\begin{figure}
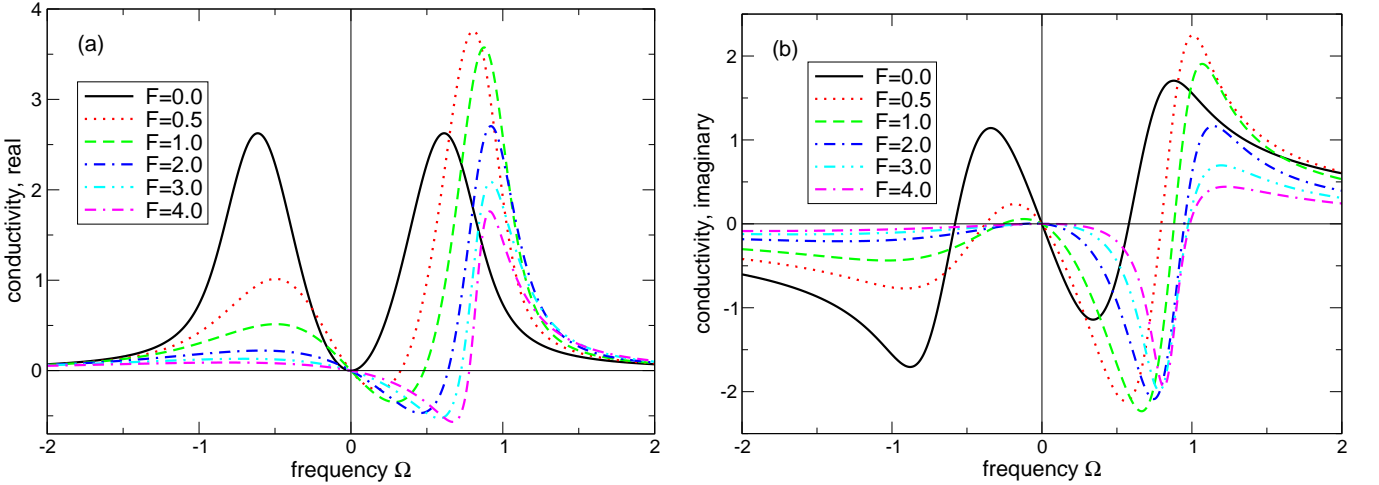

\includegraphics[width=0.49\textwidth]{fig9a.eps}\hfill 
\includegraphics[width=0.49\textwidth]{fig9b.eps}
\caption{(a) The real and (b) imaginary parts of the dimensionless dynamic conductivity (\ref{locSloc}) of an array of graphene nanoribbons at $ \mu/\Delta_0=1$, $\Gamma=0.2$, temperature $ T/\Delta_0=0.2$, $Q=1$ and several values of the dimensionless driving dc electric field ${\cal F}$. 
\label{fig:conduct}}
\end{figure}

Using the numerical integration technique described in Appendix \ref{app:integrals} we now calculate the dynamic conductivity and the absorption coefficient of a current driven system of graphene nanoribbons at different values of the input parameters. Figure \ref{fig:conduct} shows typical spectra of the real and imaginary parts of the dimensionless conductivity (\ref{locSloc}). The parameters $ \mu/\Delta_0=1$, $\Gamma=0.2$ and $ T/\Delta_0=0.2$ are the same as in Figure \ref{fig:conductivity}, the wavevector $Q=1$. In Figure \ref{fig:conduct} we show the spectra both at negative and positive frequencies. Since we have searched for a solution in the form $\propto e^{iqx-i\omega t}$, the positive (negative) frequencies correspond to the wave running in the positive (negative) direction of the $x$-axis. Since the drift velocity of electrons is positive at $E_0<0$, at positive frequencies the wave and electrons move in the same direction, while at negative frequencies they move in the opposite directions.

\begin{figure}
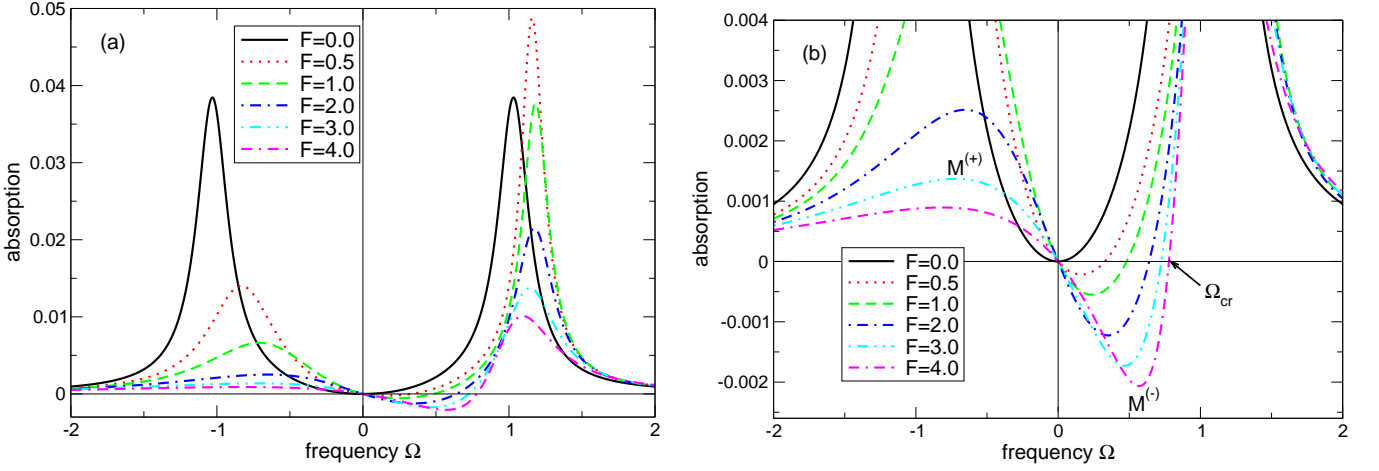

\includegraphics[width=0.49\textwidth]{fig10a.eps}\hfill 
\includegraphics[width=0.49\textwidth]{fig10b.eps}
\caption{(a) The absorption coefficient ${\cal A}$ of an array of graphene nanoribbons at $\mu/\Delta_0=1$, $\Gamma=0.2$, temperature $ T/\Delta_0=0.2$, $Q=1$ and several values of the dimensionless driving dc electric field ${\cal F}$. Other parameters are $\kappa=2.45$, $W_y/a_y=0.5$, the number of layers is 1. (b) An enlarged area of Figure (a) in the range of small and negative values of the absorption coefficient ${\cal A}$. The points $M^{(+)}=(\Omega^{(+)}_{\rm max},{\cal A}^{(+)}_{\rm max})$ and $M^{(-)}=(\Omega^{(-)}_{\rm max},{\cal A}^{(-)}_{\rm max})$ mark the positions of the maximum positive (at $\Omega<0$) and maximum negative (at $\Omega>0$) absorption. 
\label{fig:Abs1}}
\end{figure}

As seen from Figure \ref{fig:conduct}, at ${\cal F}=0$ (black solid curves) the real (imaginary) part of the dynamic conductivity is an even (odd) function of the frequency $\Omega$. The real part has two symmetric maxima at $|\Omega|\simeq 0.62$ which means that the waves running in opposite directions are absorbed equally and that the maximum absorption is the case for the wave running with the phase velocity $\omega/q\simeq 0.62 v_F$. Similar features can be seen in the absorption spectra, Figure \ref{fig:Abs1}, but the maxima are the case at $|\Omega|/Q\simeq 1$ and the absorption lines are a bit narrower, which is due to the influence of the dielectric function (\ref{dielfuncDriven}) in the denominator of Eq. (\ref{absorption}). 

When the dc field is switched on, ${\cal F}>0$, the position and the linewidth of the conductivity and absorption resonances are modified. If $\Omega<0$ (the wave and the electron beam propagate in opposite directions) the resonance is shifted to the lower frequencies and gets broader. At positive frequencies (the wave and the electron beam propagate in the same direction) the resonance is first shifted to a larger $\Omega$ and gets slightly narrower (at ${\cal F}\lesssim 2$); at stronger dc fields (at ${\cal F}\gtrsim 2$) it broadens again. The most interesting feature of the conductivity and absorption spectra is the appearance of a finite frequency interval where the real part of the conductivity and of the absorption coefficient becomes negative, see details in Figure \ref{fig:Abs1}(b). The negative absorption means that the wave is amplified taking the energy from the electron beam. This effect is the case only if the wave and electrons run in the same direction ($\Omega>0$) and if the frequency satisfies a critical condition $\Omega<\Omega_{\rm cr}$, where $\Omega_{\rm cr}({\cal F})$ depends on the driving dc electric field and other system parameters. Another interesting point $M^{(-)}({\cal F})=(\Omega^{(-)}_{\rm max},{\cal A}^{(-)}_{\rm max})$ characterizes the position of the maximum negative absorption, see Figure \ref{fig:Abs1}(b). 

At $Q=1$ and ${\cal F}=4$ the critical frequency $\Omega_{\rm cr}({\cal F}=4)$ is about 0.78 and the maximum negative absorption is about ${\cal A}^{(-)}_{\rm max}({\cal F}=4)\approx 0.2$\% at $\Omega^{(-)}_{\rm max}({\cal F}=4)\approx 0.58$. At the physical parameters outlined in Section \ref{sec:param} the values of $\Omega=0.78$ and $0.58$ correspond to 480 and 360 GHz respectively. The amplification coefficient 0.2\% does not seem to be small if to remember that we are dealing with only a monolayer of atoms with the very low areal density of electrons $\simeq 2.3\times 10^8$ cm$^{-2}$ (at parameters from Section \ref{sec:param}, see the estimate in Eq. (\ref{dens-estimate})). 

\begin{figure}
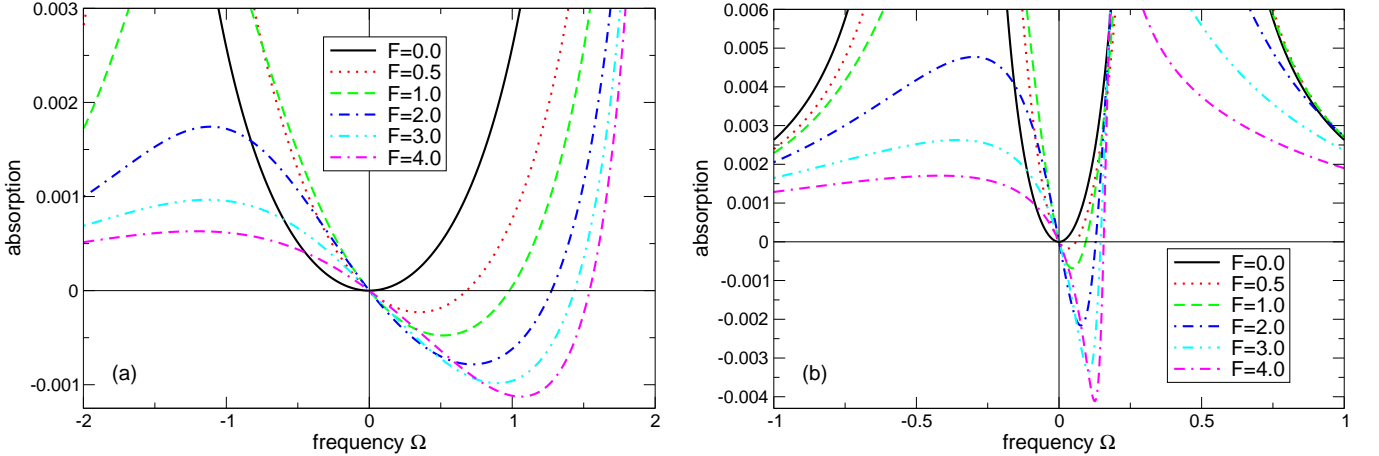

\includegraphics[width=0.49\textwidth]{fig11a.eps}\hfill 
\includegraphics[width=0.49\textwidth]{fig11b.eps}
\caption{The influence of the wavevector $Q$: The absorption coefficient ${\cal A}$ of an array of graphene nanoribbons at $\mu/\Delta_0=1$, $\Gamma=0.2$, temperature $ T/\Delta_0=0.2$, several values of the dimensionless driving dc electric field ${\cal F}$ and (a) $Q=2$ and (b) $Q=0.2$. Other parameters are $\kappa=2.45$, $W_y/a_y=0.5$, the number of layers is 1. 
\label{fig:Abs2}}
\end{figure}

If the wavevector $Q$ is two times larger, Figure \ref{fig:Abs2}(a), the critical frequency $\Omega_{\rm cr}({\cal F}=4)$ increases up to $\Omega_{\rm cr}({\cal F}=4)\approx 1.54$ (corresponds to $\approx 0.95$ THz) but the maximum amplification (negative absorption) decreases, ${\cal A}^{(-)}_{\rm max}({\cal F}=4)\approx 0.11$\%. If the wavevector gets smaller, see Figure \ref{fig:Abs2}(b) for $Q=0.2$, the value of ${\cal A}^{(-)}_{\rm max}({\cal F}=4)\approx 0.41$\% increases, but the critical frequency $\Omega_{\rm cr}({\cal F}=4)\approx 0.156$ (corresponds to $\approx 97$ GHz) decreases.

\begin{figure}
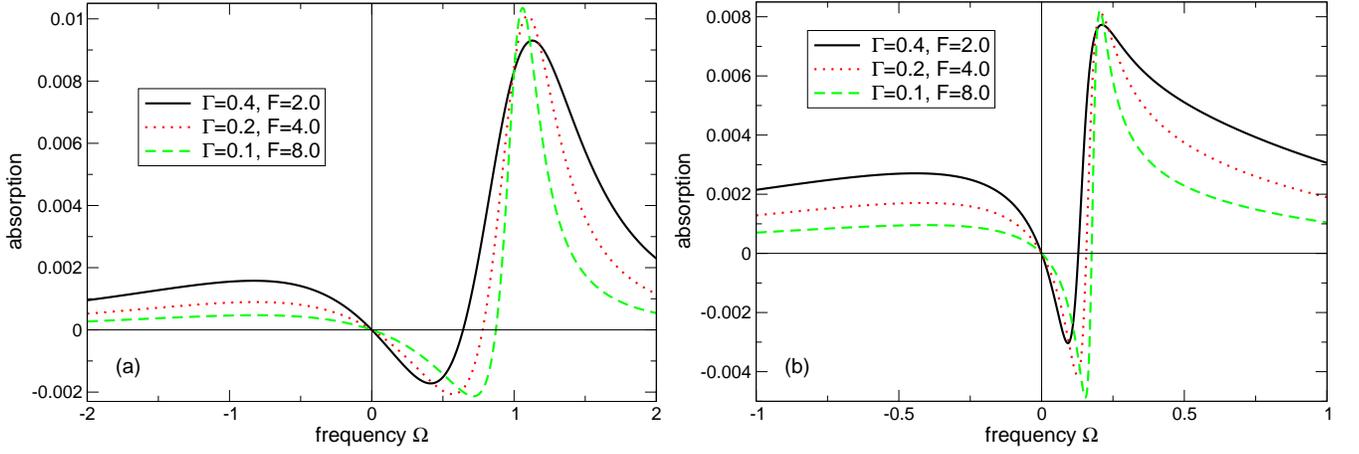

\includegraphics[width=0.49\textwidth]{fig12a.eps}
\includegraphics[width=0.49\textwidth]{fig12b.eps}
\caption{The influence of the scattering rate $\gamma$: The absorption coefficient ${\cal A}$ of an array of graphene nanoribbons at (a) $Q=1$ and (b) $Q=0.2$; other parameters are $ \mu/\Delta_0=1$, $ T/\Delta_0=0.2$, $\kappa=2.45$, and $W_y/a_y=0.5$. 
\label{fig:Abs3}}
\end{figure}

Figure \ref{fig:Abs3} illustrates the influence of the scattering rate $\gamma$ on the absorption spectra. Here we show the spectra for the ${\cal F}=4$ curves from Figure \ref{fig:Abs1}(b) ($Q=1$) and \ref{fig:Abs2}(b) ($Q=0.2$) together with two other curves corresponding to two times smaller and two times larger values of the scattering rate $\gamma$. We notice that the scattering rate $\gamma$ enters two dimensionless quantities $\Gamma$ and ${\cal F}$, see Eq. (\ref{param}). Therefore reducing $\gamma$ by a factor of two leads to the reduction of $\Gamma$ by the factor of two and to the increase of ${\cal F}$ by the same factor. One sees that increasing $\gamma$ reduces both the negative absorption region ($\Omega_{\rm cr}$ decreases) and the maximum amplification value ${\cal A}^{(-)}_{\rm max}$.

For numerical parameters used in this paper (see Section \ref{sec:param}) the value of $\Gamma\simeq 0.2$ corresponds to the mean free path $l\simeq 1.3$ $\mu$m. The $l$-values of this (and larger) scale have been observed in graphene sandwiched between two \textit{h}-BN crystals, e.g. Ref.  \cite{Mayorov11b}. It should be noticed that the scattering rate $\gamma$ should be considered as a phenomenological parameter which may depend, for example, on the electron density, temperature or dc electric field. For example, at ${\cal F}\gg 1$ electrons may get enough energy to emit optical phonons or to be scattered to higher electron subbands. The first process is however unlikely since the typical energy of the optical phonons in the considered systems ($\sim 50-100$ meV) is $20-40$ times larger than the energy scale in our problem ($\Delta_0\simeq 2.5$ meV, Section \ref{sec:param}). The scattering to the higher subbands is more likely but we believe that such a scattering should not dramatically influence our results: the probability of the elastic (e.g. impurity) scattering from $|1,p_+\rangle$ to $|k,p_+\rangle$ states is much higher than to the $|k,p_-\rangle$ states (due to the smaller momentum transfer) but the processes $|1,p_+\rangle\to |k,p_+\rangle$ lead to small current changes since after the scattering electrons continue to move in the same direction (here the first number in the ket states denotes the subband index, $k>1$, and the $|k,p_\pm\rangle$ states are the states with the same energy but different, positive and negative, momentum).  

\begin{figure}
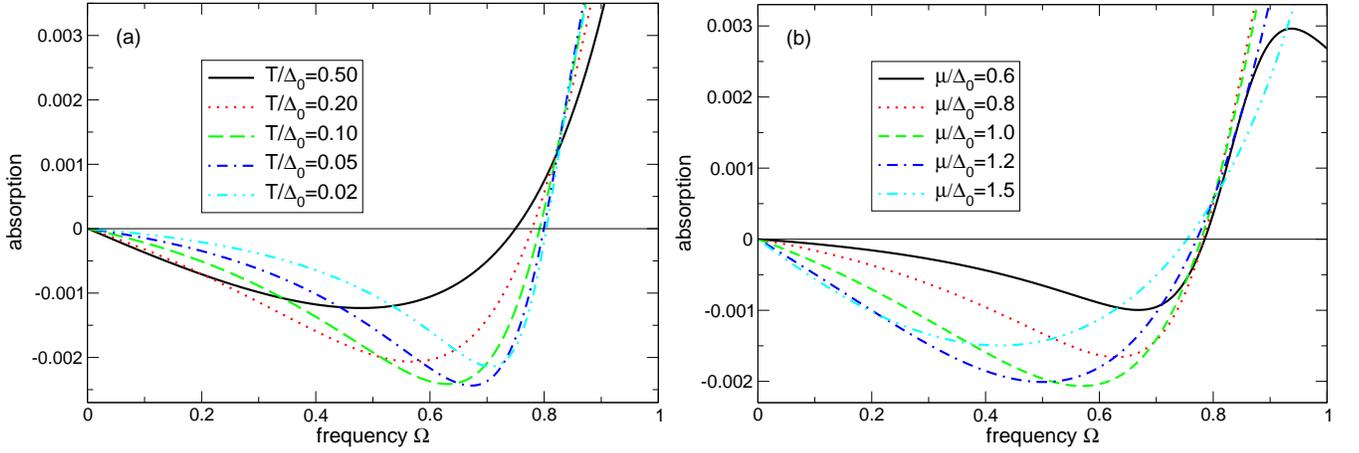

\includegraphics[width=0.49\textwidth]{fig13a.eps}
\includegraphics[width=0.49\textwidth]{fig13b.eps}
\caption{The influence of the temperature and the chemical potential: The absorption coefficient ${\cal A}$ of an array of graphene nanoribbons at $\Gamma=0.2$, $Q=1$, ${\cal F}=4$ and (a) different temperatures at $\mu/\Delta_0=1$ and (b) different chemical potentials $\tilde\mu$ at $\tilde T=0.2$. Other parameters are $\kappa=2.45$, $W_y/a_y=0.5$, the number of layers is 1. 
\label{fig:AbsTempMu}}
\end{figure}

The influence of the temperature and the chemical potential on the maximum value of the negative absorption is illustrated in Figure \ref{fig:AbsTempMu}. One sees that the critical point $\Omega_{\rm cr}$ slightly grows with decreasing temperature and the chemical potential (i.e. with decreasing the electron density). The maximum amplification value ${\cal A}^{(-)}_{\rm max}$ has an optimum as a function of both $\tilde T$ and $\tilde \mu$. At the parameters of Figure \ref{fig:AbsTempMu} this optimum lies at approximately $0.24$\% and is achieved at $\tilde \mu\approx 1$ and $\tilde T\approx 0.05-0.2$.

The rather small value of the parameter $\tilde T\simeq 0.2$ corresponds, in the chosen numerical example with $W_y=400$ nm and $\Delta_0\approx 28$ K (Section \ref{sec:form_prob}), to cryogenic temperatures $T\simeq 5.6$ K. The operation temperature of devices based on the discussed effect (e.g. of the Smith-Purcell-type emitter of terahertz radiation,  see discussion in Section \ref{sec:conclusions}) can be increased if the nanoribbons could be made substantially narrower. For example, at $W_y=10$ nm the gap corresponds to $\Delta_0\simeq 1120$ K, so that the room-temperature operation becomes feasible. 

Analyzing all so far presented results one can notice that the critical frequency $\Omega_{\rm cr}$, which restricts the negative absorption region from above, is related to the average velocity of electrons in our system of graphene nanoribbons, Figure \ref{fig:aveveloc}. Comparing the absorption plots with Figure \ref{fig:aveveloc} one can observe that $\Omega_{\rm cr}\approx Q V_{\rm dr}$. The amplification of the electromagnetic wave is thus the case when the drift velocity of electrons exceeds the phase velocity of the wave,
\be 
\omega/q\lesssim \bar v_{\rm dr}.\label{condition}
\ee
Under this condition the electron flow transmits its energy to the wave and amplifies it; under the opposite condition the wave transmits its energy to the electron system which leads to an additional damping of the wave and to the electron drag effect (not studied here). Notice that being written in the form (\ref{condition}) the amplification condition does not depend on details of the considered structure and its parameters and has therefore a larger range of applicability than the initial model. 

The studied phenomenon is closely related to the acoustoelectric effect, in which the running electric-field excitation is created by a surface acoustic wave propagating in a piezoelectric material underlying the 2D electron system. The acoustoelectric interaction has been widely used and studied in semiconductor structures with the 2D electron gas, see e.g. \cite{Wixforth86,Falko93}, as well as recently in graphene and other (e.g. MoS$_2$) two-dimensional crystals \cite{Bandhu13,Bandhu14,Preciado15}. It should be noticed however that the discussed effect in graphene offers substantially more opportunities as compared to semiconductors: while in semiconductors the drift velocity cannot be made very high so that one can only study the interaction of the drifting electrons with the (relatively slow) acoustic waves ($\omega/q\simeq 10^5$ cm/s), in graphene the drift velocity can approach the Fermi velocity $\simeq 10^8$ cm/s, which makes it possible to study the interaction of drifting electrons with faster electromagnetic excitations, e.g. with 2D plasma waves. This work thus also contributes to the theory of graphene-based voltage-tunable terahertz emitter proposed and discussed in Refs. \cite{Mikhailov13a,Moskalenko14}.

\begin{figure}
\includegraphics[width=0.49\textwidth]{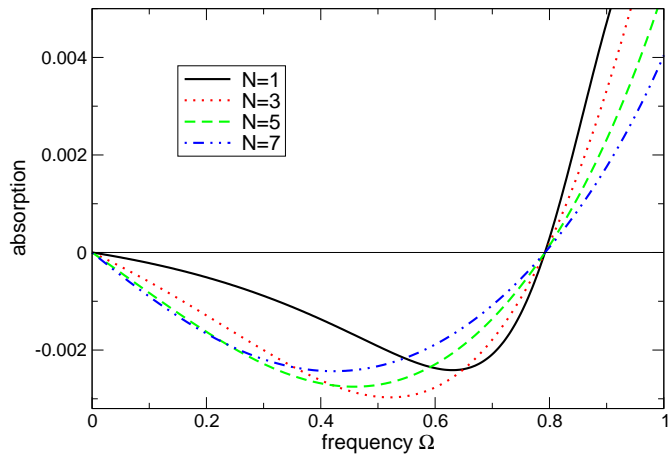}
\caption{The influence of the number of layers: The absorption coefficient ${\cal A}$ of an array of graphene nanoribbons at  $\mu/\Delta_0=1$, $\Gamma=0.2$, $T/\Delta_0=0.2$, $Q=1$, $\kappa=2.45$, $W_y/a_y=0.5$ at different numbers $N$ of the graphene-nanoribbons layers.
\label{fig:Abs5}}
\end{figure}

Finally, we would like to discuss the following point. The absolute value of the amplification factor ${\cal A}^{(-)}_{\rm max}$ was found to be rather small, below $\sim 0.5$\%, which is mainly due to the low density of electrons in a system which has only one (structured) monolayer of carbon atoms. Naturally the question arises, whether the factor ${\cal A}^{(-)}_{\rm max}$ can be increased by forming a many-layered system of parallel graphene-nanoribbon arrays. Figure \ref{fig:Abs5} answers this question. It shows the absorption spectra of an array of the current-driven graphene nanoribbons for different numbers of layers $N$. At $N=1$ the maximum negative absorption ${\cal A}^{(-)}_{\rm max}$ is about $0.24$\%. When $N$ increases, the absolute value of ${\cal A}^{(-)}_{\rm max}$ first grows with $N$, indeed, but at $N\simeq 3$ it reaches a maximum ($\simeq 0.3$\%) and then decreases again. Such a behavior becomes clear from Eqs. (\ref{absorption}), (\ref{sigma2D}) and (\ref{dielfuncDriven}): The number of layers $N$ enters both the nominator ($\sim N$) and the denominator (which grows as $\sim N^2$ at $N\to \infty$). The suppression of the factor ${\cal A}^{(-)}_{\rm max}$ at large $N$ is thus due to the dielectric function (\ref{dielfuncDriven}). This suppression is unavoidable but may depend on (and be controlled by) the specific dielectric environment. By a special design of the environment one can probably increase the absolute value of the amplification factor ${\cal A}^{(-)}_{\rm max}$ but this problem is not considered in the present work.

\section{Conclusions\label{sec:conclusions}}

To summarize, we have studied the wave-vector, frequency, dc electric field, scattering rate, electron density, temperature and the number-of-layers dependence of the dynamic conductivity, dielectric function and the absorption coefficient in a direct current driven array of graphene nanoribbons. The influence of the dc electric field is taken into account non-perturbatively, the influence of the ac field -- within the linear response theory, the scattering processes -- within the simple relaxation time approximation. We have shown that, at frequencies satisfying the condition (\ref{condition}), the propagating electromagnetic wave can be amplified at the expense of the energy of the direct current source. We have analyzed the optimal conditions for the wave amplification depending on all system parameters. 

We have assumed that the external electric field acting on the system is given by the formula (\ref{extfield}) but did not discuss how the ac part of this field is created. This can be done by different means. For example, the running electric-field wave can be produced by the surface acoustic waves propagating along the surface of a piezoelectric material covered by graphene \cite{Bandhu13,Bandhu14}, a semiconductor 2D crystal \cite{Preciado15}, or by a graphene nanostructure as considered in this paper. In this case the phase velocity of the wave is about $\omega/q\simeq 6\times 10^5$ cm/s ($\omega/ qv_F\simeq 6\times10^{-3}$) and the required drift velocity and the dc electric field are very low ($\bar v_{\rm dr}/ v_F\ll 1$, ${\cal F}\ll 1$), see Figure \ref{fig:aveveloc}. 

Another opportunity to create the running electric-field wave (\ref{extfield}) is to irradiate the graphene-nanoribbon array with an adjacent grating structure by electromagnetic radiation\cite{Mikhailov13a}. The grating transforms the incident wave to the electromagnetic excitation running along the 2D layer with the phase velocity $\omega/q\ll c$ (with $c$ being the speed of light), thus realizing a Smith-Purcell-type emitter of radiation \cite{Smith53}. In this case the phase velocity of the wave can be comparable with the Fermi velocity of electrons in graphene, $\omega/q\lesssim \bar v_{\rm dr}\lesssim v_F\approx 10^8$ cm/s, which make it feasible to extend the operation frequency range up to terahertz. Attempts to realize such emitters of terahertz radiation have been made in the past on the basis of semiconductor structures \cite{Tsui80,Hirakawa95}, for reviews see, e.g. \cite{Mikhailov98c,Otsuji11,Otsuji12}. Recently it has been proposed to use graphene for this purpose \cite{Mikhailov13a,Moskalenko14}. This work thus contributes to the realization of graphene-based voltage-tunable emitters of sub-terahertz and terahertz radiation. 

We thank Geoffrey Nash, J\'er\^ome Faist, Isaac Luxmoore, Federico Valmorra, Janine Keller and Hua Qin for numerous discussions of issues related to the experimental realization of the graphene-based terahertz emitters. The work has received funding from the European Union under the FET-open grant GOSFEL (grant agreement No. 296391) and the European Union's Horizon 2020 research and innovation programme Graphene Flagship under grant agreement No. 696656.

\appendix

\section{Method of calculating the integrals \label{app:integrals}}
 
We aim to develop an efficient method of a numerical evaluation of the following integrals
\be 
{\cal Z}_n(\Omega,Q,\Gamma,\tilde \mu,\tilde T,{\cal F})=-
\int_{-\infty}^\infty V^n(P)dP
\int_{-\infty}^{P} 
{\cal U}(P',\tilde\mu,\tilde T,{\cal F})
{\cal D}(P,P';\Omega,Q,\Gamma,{\cal F})
dP' ,\label{funcZ}
\ee
where $n=0,1$, ${\cal D}(P,P';\Omega,Q,\Gamma,{\cal F})$ is the exponential function (\ref{funcD-exponent}) and $V(P)$ is the dimensionless velocity (\ref{dimlessvelocity}). The functions ${\cal K}_n$ and ${\cal L}_n$ in Eqs. (\ref{funcKn})--(\ref{funcLn}) are special cases of (\ref{funcZ}) if 
\be 
{\cal U}(P,\tilde\mu,\tilde T,{\cal F})=\frac{df_0(P,\tilde\mu,\tilde T,{\cal F})}{dP}=\frac{f_0(P,\tilde\mu,\tilde T,{\cal F})
-f_{\rm eq}(P,\tilde\mu,\tilde T) }{{\cal F}},
\ee
and 
\be 
{\cal U}(P,\tilde\mu,\tilde T,{\cal F})=\left(\cosh\frac{\sqrt{1+P^2}-\tilde\mu}{2\tilde T}\right)^{-2},
\ee
respectively; here 
\be 
f_{\rm eq}(P,\tilde\mu,\tilde T) =\frac{1}{1+\exp\left(\frac{\sqrt{1+P^2}-\tilde\mu }{\tilde T }\right)} 
\ee
is the equilibrium Fermi distribution function expressed in the dimensionless variables. 
A direct numerical integration in Eq. (\ref{funcZ}) runs into problems: the function ${\cal D}(P,P')$ strongly oscillates, which leads to a very long integration time as well as to error messages like ``the required accuracy was not achieved''. To solve this problem we had to find a more elegant way to calculate the integral (\ref{funcZ}).

First, let us change the order of integration over $dP$ and $dP'$ and make the substitution $P'=\sinh A$, $P=\sinh X$. This gives 
\be 
{\cal Z}_n(\Omega,Q,\Gamma,\tilde \mu,\tilde T,{\cal F})=
\int_{-\infty}^\infty 
{\cal U}(\sinh A,\tilde\mu,\tilde T,{\cal F})
{\cal V}_n(\Omega,Q,\Gamma,{\cal F},A)\cosh A  dA ,\label{eqo}
\ee
where the new function ${\cal V}_n$ is defined as 
\be 
{\cal V}_n(\Omega,Q,\Gamma,{\cal F},A)=-
\int_A^{\infty} \tanh^n X \cosh X dX
\exp\left(i\frac{(\Omega+i\Gamma)(\sinh X-\sinh A)-Q(\cosh X-\cosh A)}{{\cal F}\Gamma}\right).\label{funcVn}
\ee
The function ${\cal U}(\sinh A,\tilde\mu,\tilde T,{\cal F})$ in Eq. (\ref{eqo}) is smooth and localized, therefore the problem is only in the functions ${\cal V}_n(\Omega,Q,\Gamma,{\cal F},A)$, Eq. (\ref{funcVn}). The integrand in (\ref{funcVn}) is a strongly oscillating function of $X$. But, considering it as a function of a complex variable $Z$,
\be 
\tanh^n Z \cosh Z \exp\left(i\frac{(\Omega+i\Gamma)(\sinh Z-\sinh A)-Q(\cosh Z-\cosh A)}{{\cal F}\Gamma}\right)\label{integrand}
\ee
(here $n=0$ and $1$), we see that it is an analytical function in the whole complex plane $Z$. Therefore we can choose another integration path from the point $Z=A$ to the complex infinity without changing the value of the integral. Choosing the integration path so that the integrand oscillations are substantially suppressed we can calculate the integrals ${\cal V}_n(\Omega,Q,\Gamma,{\cal F},A)$ much faster and without error messages. 

\begin{figure}
\includegraphics[width=0.49\textwidth]{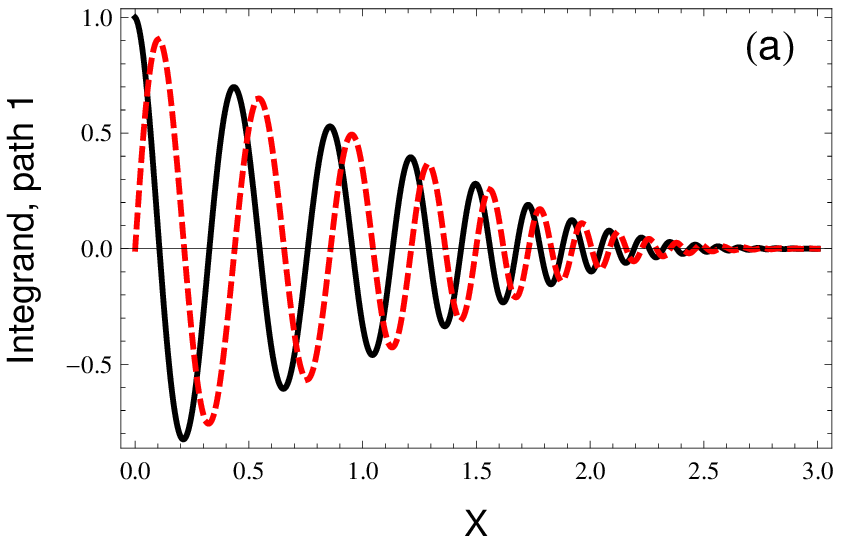}\\
\includegraphics[width=0.48\textwidth]{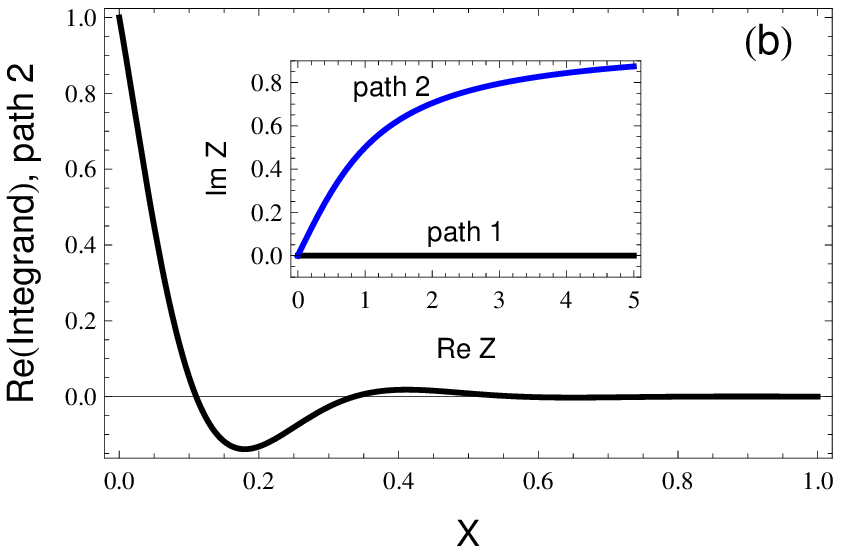}
\includegraphics[width=0.51\textwidth]{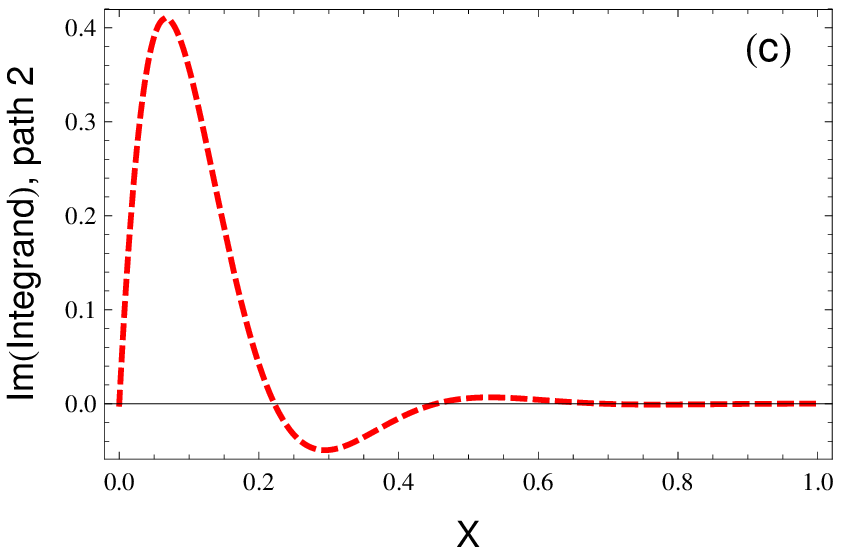}
\caption{(a) The real (solid black curve) and imaginary (red dashed curve) parts of the integrand (\ref{integrand}) if the complex variable $Z$ varies along the real axis [path 1: $Z=X$, $X:=(A\to \infty)$]; (b) the real and (c) imaginary parts of the integrand (\ref{integrand}) if $Z$ varies along the path 2: $Z=X+iY(X)$, $X:=(A\to \infty)$, where $Y(X)$ is chosen so that the oscillations are minimized. Inset in (b) shows both paths. The parameters used in these plots are: $n= 0$, $\Omega= 3$, $Q= 1$, $\Gamma= 0.2$, ${\cal F}= 1$, $A= 0$.
\label{fig:path12}}
\end{figure}

Figure \ref{fig:path12} illustrates this approach. In Figure \ref{fig:path12}(a) we show the real and imaginary parts of the integrand (\ref{integrand}) if $Z$ varies along the integration path 1 (the real axis $Z=X$); the path 1 is shown by the black line in the inset to Figure \ref{fig:path12}(b). One sees that both the real and imaginary parts weakly decay and strongly oscillate when $X$ increases. In Figures \ref{fig:path12}(b,c) we choose another path, $Z(X)=X+iY(X)$, $A<X<\infty$, shown by the blue curve in the inset to Figure \ref{fig:path12}(b). The real and imaginary parts of the integrand (\ref{integrand}) (shown in Figures \ref{fig:path12}(b) and (c) respectively) tend to zero much faster and without oscillations. The path 2 may depend on parameters $\Omega$, $Q$, $\Gamma$ and $A$ but a suitable path can always be found.

\begin{thebibliography}{49}
\expandafter\ifx\csname natexlab\endcsname\relax\def\natexlab#1{#1}\fi
\expandafter\ifx\csname bibnamefont\endcsname\relax
  \def\bibnamefont#1{#1}\fi
\expandafter\ifx\csname bibfnamefont\endcsname\relax
  \def\bibfnamefont#1{#1}\fi
\expandafter\ifx\csname citenamefont\endcsname\relax
  \def\citenamefont#1{#1}\fi
\expandafter\ifx\csname url\endcsname\relax
  \def\url#1{\texttt{#1}}\fi
\expandafter\ifx\csname urlprefix\endcsname\relax\def\urlprefix{URL }\fi
\providecommand{\bibinfo}[2]{#2}
\providecommand{\eprint}[2][]{\url{#2}}

\bibitem[{\citenamefont{Mikhailov}(2007)}]{Mikhailov07e}
\bibinfo{author}{\bibfnamefont{S.~A.} \bibnamefont{Mikhailov}},
  \bibinfo{journal}{Europhys. Lett.} \textbf{\bibinfo{volume}{79}},
  \bibinfo{pages}{27002} (\bibinfo{year}{2007}).

\bibitem[{\citenamefont{Dragoman et~al.}(2010)\citenamefont{Dragoman, Neculoiu,
  Deligeorgis, Konstantinidis, Dragoman, Cismaru, Muller, and
  Plana}}]{Dragoman10}
\bibinfo{author}{\bibfnamefont{M.}~\bibnamefont{Dragoman}},
  \bibinfo{author}{\bibfnamefont{D.}~\bibnamefont{Neculoiu}},
  \bibinfo{author}{\bibfnamefont{G.}~\bibnamefont{Deligeorgis}},
  \bibinfo{author}{\bibfnamefont{G.}~\bibnamefont{Konstantinidis}},
  \bibinfo{author}{\bibfnamefont{D.}~\bibnamefont{Dragoman}},
  \bibinfo{author}{\bibfnamefont{A.}~\bibnamefont{Cismaru}},
  \bibinfo{author}{\bibfnamefont{A.~A.} \bibnamefont{Muller}},
  \bibnamefont{and} \bibinfo{author}{\bibfnamefont{R.}~\bibnamefont{Plana}},
  \bibinfo{journal}{Appl. Phys. Lett.} \textbf{\bibinfo{volume}{97}},
  \bibinfo{pages}{093101} (\bibinfo{year}{2010}).

\bibitem[{\citenamefont{Hotopan et~al.}(2011)\citenamefont{Hotopan, {Ver
  Hoeye}, Vazquez, Camblor, Fern\'andez, Las~Heras, \'Alvarez, and
  Men\'endez}}]{Hotopan11}
\bibinfo{author}{\bibfnamefont{G.}~\bibnamefont{Hotopan}},
  \bibinfo{author}{\bibfnamefont{S.}~\bibnamefont{{Ver Hoeye}}},
  \bibinfo{author}{\bibfnamefont{C.}~\bibnamefont{Vazquez}},
  \bibinfo{author}{\bibfnamefont{R.}~\bibnamefont{Camblor}},
  \bibinfo{author}{\bibfnamefont{M.}~\bibnamefont{Fern\'andez}},
  \bibinfo{author}{\bibfnamefont{F.}~\bibnamefont{Las~Heras}},
  \bibinfo{author}{\bibfnamefont{P.}~\bibnamefont{\'Alvarez}},
  \bibnamefont{and}
  \bibinfo{author}{\bibfnamefont{R.}~\bibnamefont{Men\'endez}},
  \bibinfo{journal}{Progress In Electromagnetic Research}
  \textbf{\bibinfo{volume}{118}}, \bibinfo{pages}{57} (\bibinfo{year}{2011}).

\bibitem[{\citenamefont{Hendry et~al.}(2010)\citenamefont{Hendry, Hale, Moger,
  Savchenko, and Mikhailov}}]{Hendry10}
\bibinfo{author}{\bibfnamefont{E.}~\bibnamefont{Hendry}},
  \bibinfo{author}{\bibfnamefont{P.~J.} \bibnamefont{Hale}},
  \bibinfo{author}{\bibfnamefont{J.~J.} \bibnamefont{Moger}},
  \bibinfo{author}{\bibfnamefont{A.~K.} \bibnamefont{Savchenko}},
  \bibnamefont{and} \bibinfo{author}{\bibfnamefont{S.~A.}
  \bibnamefont{Mikhailov}}, \bibinfo{journal}{Phys. Rev. Lett.}
  \textbf{\bibinfo{volume}{105}}, \bibinfo{pages}{097401}
  (\bibinfo{year}{2010}).

\bibitem[{\citenamefont{Wu et~al.}(2011)\citenamefont{Wu, Zhang, Yan, Bian,
  Wang, Bai, Lu, Zhao, and Wang}}]{Wu11}
\bibinfo{author}{\bibfnamefont{R.}~\bibnamefont{Wu}},
  \bibinfo{author}{\bibfnamefont{Y.}~\bibnamefont{Zhang}},
  \bibinfo{author}{\bibfnamefont{S.}~\bibnamefont{Yan}},
  \bibinfo{author}{\bibfnamefont{F.}~\bibnamefont{Bian}},
  \bibinfo{author}{\bibfnamefont{W.}~\bibnamefont{Wang}},
  \bibinfo{author}{\bibfnamefont{X.}~\bibnamefont{Bai}},
  \bibinfo{author}{\bibfnamefont{X.}~\bibnamefont{Lu}},
  \bibinfo{author}{\bibfnamefont{J.}~\bibnamefont{Zhao}}, \bibnamefont{and}
  \bibinfo{author}{\bibfnamefont{E.}~\bibnamefont{Wang}},
  \bibinfo{journal}{Nano Lett.} \textbf{\bibinfo{volume}{11}},
  \bibinfo{pages}{5159 } (\bibinfo{year}{2011}).

\bibitem[{\citenamefont{Gu et~al.}(2012)\citenamefont{Gu, Petrone, McMillan,
  {van der Zande}, Yu, Lo, Kwong, Hone, and Wong}}]{Gu12}
\bibinfo{author}{\bibfnamefont{T.}~\bibnamefont{Gu}},
  \bibinfo{author}{\bibfnamefont{N.}~\bibnamefont{Petrone}},
  \bibinfo{author}{\bibfnamefont{J.~F.} \bibnamefont{McMillan}},
  \bibinfo{author}{\bibfnamefont{A.}~\bibnamefont{{van der Zande}}},
  \bibinfo{author}{\bibfnamefont{M.}~\bibnamefont{Yu}},
  \bibinfo{author}{\bibfnamefont{G.~Q.} \bibnamefont{Lo}},
  \bibinfo{author}{\bibfnamefont{D.~L.} \bibnamefont{Kwong}},
  \bibinfo{author}{\bibfnamefont{J.}~\bibnamefont{Hone}}, \bibnamefont{and}
  \bibinfo{author}{\bibfnamefont{C.~W.} \bibnamefont{Wong}},
  \bibinfo{journal}{Nature Photonics} \textbf{\bibinfo{volume}{6}},
  \bibinfo{pages}{554} (\bibinfo{year}{2012}).

\bibitem[{\citenamefont{Zhang et~al.}(2012)\citenamefont{Zhang, Virally, Bao,
  Ping, Massar, Godbout, and Kockaert}}]{Zhang12}
\bibinfo{author}{\bibfnamefont{H.}~\bibnamefont{Zhang}},
  \bibinfo{author}{\bibfnamefont{S.}~\bibnamefont{Virally}},
  \bibinfo{author}{\bibfnamefont{Q.}~\bibnamefont{Bao}},
  \bibinfo{author}{\bibfnamefont{L.~K.} \bibnamefont{Ping}},
  \bibinfo{author}{\bibfnamefont{S.}~\bibnamefont{Massar}},
  \bibinfo{author}{\bibfnamefont{N.}~\bibnamefont{Godbout}}, \bibnamefont{and}
  \bibinfo{author}{\bibfnamefont{P.}~\bibnamefont{Kockaert}},
  \bibinfo{journal}{Optics Letters} \textbf{\bibinfo{volume}{37}},
  \bibinfo{pages}{1856} (\bibinfo{year}{2012}).

\bibitem[{\citenamefont{Bykov et~al.}(2012)\citenamefont{Bykov, Murzina, Rybin,
  and Obraztsova}}]{Bykov12}
\bibinfo{author}{\bibfnamefont{A.~Y.} \bibnamefont{Bykov}},
  \bibinfo{author}{\bibfnamefont{T.~V.} \bibnamefont{Murzina}},
  \bibinfo{author}{\bibfnamefont{M.~G.} \bibnamefont{Rybin}}, \bibnamefont{and}
  \bibinfo{author}{\bibfnamefont{E.~D.} \bibnamefont{Obraztsova}},
  \bibinfo{journal}{Phys. Rev. B} \textbf{\bibinfo{volume}{85}},
  \bibinfo{pages}{121413(R)} (\bibinfo{year}{2012}).

\bibitem[{\citenamefont{Kumar et~al.}(2013)\citenamefont{Kumar, Kumar,
  Gerstenkorn, Wang, Chiu, Smirl, and Zhao}}]{Kumar13}
\bibinfo{author}{\bibfnamefont{N.}~\bibnamefont{Kumar}},
  \bibinfo{author}{\bibfnamefont{J.}~\bibnamefont{Kumar}},
  \bibinfo{author}{\bibfnamefont{C.}~\bibnamefont{Gerstenkorn}},
  \bibinfo{author}{\bibfnamefont{R.}~\bibnamefont{Wang}},
  \bibinfo{author}{\bibfnamefont{H.-Y.} \bibnamefont{Chiu}},
  \bibinfo{author}{\bibfnamefont{A.~L.} \bibnamefont{Smirl}}, \bibnamefont{and}
  \bibinfo{author}{\bibfnamefont{H.}~\bibnamefont{Zhao}},
  \bibinfo{journal}{Phys. Rev. B} \textbf{\bibinfo{volume}{87}},
  \bibinfo{pages}{121406(R)} (\bibinfo{year}{2013}).

\bibitem[{\citenamefont{Hong et~al.}(2013)\citenamefont{Hong, Dadap, Petrone,
  Yeh, Hone, and {Osgood, Jr.}}}]{Hong13}
\bibinfo{author}{\bibfnamefont{S.-Y.} \bibnamefont{Hong}},
  \bibinfo{author}{\bibfnamefont{J.~I.} \bibnamefont{Dadap}},
  \bibinfo{author}{\bibfnamefont{N.}~\bibnamefont{Petrone}},
  \bibinfo{author}{\bibfnamefont{P.-C.} \bibnamefont{Yeh}},
  \bibinfo{author}{\bibfnamefont{J.}~\bibnamefont{Hone}}, \bibnamefont{and}
  \bibinfo{author}{\bibfnamefont{R.~M.} \bibnamefont{{Osgood, Jr.}}},
  \bibinfo{journal}{Phys. Rev. X} \textbf{\bibinfo{volume}{3}},
  \bibinfo{pages}{021014} (\bibinfo{year}{2013}).

\bibitem[{\citenamefont{An et~al.}(2014)\citenamefont{An, Rowe, Dougherty, Lee,
  and Diebold}}]{An14}
\bibinfo{author}{\bibfnamefont{Y.~Q.} \bibnamefont{An}},
  \bibinfo{author}{\bibfnamefont{J.~E.} \bibnamefont{Rowe}},
  \bibinfo{author}{\bibfnamefont{D.~B.} \bibnamefont{Dougherty}},
  \bibinfo{author}{\bibfnamefont{J.~U.} \bibnamefont{Lee}}, \bibnamefont{and}
  \bibinfo{author}{\bibfnamefont{A.~C.} \bibnamefont{Diebold}},
  \bibinfo{journal}{Phys. Rev. B} \textbf{\bibinfo{volume}{89}},
  \bibinfo{pages}{115310} (\bibinfo{year}{2014}).

\bibitem[{\citenamefont{Mikhailov}(2011)}]{Mikhailov11c}
\bibinfo{author}{\bibfnamefont{S.~A.} \bibnamefont{Mikhailov}},
  \bibinfo{journal}{Phys. Rev. B} \textbf{\bibinfo{volume}{84}},
  \bibinfo{pages}{045432} (\bibinfo{year}{2011}).

\bibitem[{\citenamefont{Yao and Belyanin}(2013)}]{Yao13}
\bibinfo{author}{\bibfnamefont{X.~H.} \bibnamefont{Yao}} \bibnamefont{and}
  \bibinfo{author}{\bibfnamefont{A.}~\bibnamefont{Belyanin}},
  \bibinfo{journal}{J. Phys. Condens. Matter} \textbf{\bibinfo{volume}{25}},
  \bibinfo{pages}{054203} (\bibinfo{year}{2013}).

\bibitem[{\citenamefont{Cheng et~al.}(2014{\natexlab{a}})\citenamefont{Cheng,
  Vermeulen, and Sipe}}]{Cheng14a}
\bibinfo{author}{\bibfnamefont{J.~L.} \bibnamefont{Cheng}},
  \bibinfo{author}{\bibfnamefont{N.}~\bibnamefont{Vermeulen}},
  \bibnamefont{and} \bibinfo{author}{\bibfnamefont{J.~E.} \bibnamefont{Sipe}},
  \bibinfo{journal}{New J. Phys.} \textbf{\bibinfo{volume}{16}},
  \bibinfo{pages}{053014} (\bibinfo{year}{2014}{\natexlab{a}}).

\bibitem[{\citenamefont{Cheng et~al.}(2014{\natexlab{b}})\citenamefont{Cheng,
  Vermeulen, and Sipe}}]{Cheng14b}
\bibinfo{author}{\bibfnamefont{J.~L.} \bibnamefont{Cheng}},
  \bibinfo{author}{\bibfnamefont{N.}~\bibnamefont{Vermeulen}},
  \bibnamefont{and} \bibinfo{author}{\bibfnamefont{J.~E.} \bibnamefont{Sipe}},
  \bibinfo{journal}{Optics Express} \textbf{\bibinfo{volume}{22}},
  \bibinfo{pages}{15868} (\bibinfo{year}{2014}{\natexlab{b}}).

\bibitem[{\citenamefont{Peres et~al.}(2014)\citenamefont{Peres, Bludov, Santos,
  Jauho, and Vasilevskiy}}]{Peres14}
\bibinfo{author}{\bibfnamefont{N.~M.~R.} \bibnamefont{Peres}},
  \bibinfo{author}{\bibfnamefont{Y.~V.} \bibnamefont{Bludov}},
  \bibinfo{author}{\bibfnamefont{J.~E.} \bibnamefont{Santos}},
  \bibinfo{author}{\bibfnamefont{A.-P.} \bibnamefont{Jauho}}, \bibnamefont{and}
  \bibinfo{author}{\bibfnamefont{M.~I.} \bibnamefont{Vasilevskiy}},
  \bibinfo{journal}{Phys. Rev. B} \textbf{\bibinfo{volume}{90}},
  \bibinfo{pages}{125425} (\bibinfo{year}{2014}).

\bibitem[{\citenamefont{Smirnova et~al.}(2014)\citenamefont{Smirnova,
  Shadrivov, Miroshnichenko, Smirnov, and Kivshar}}]{Smirnova14}
\bibinfo{author}{\bibfnamefont{D.~A.} \bibnamefont{Smirnova}},
  \bibinfo{author}{\bibfnamefont{I.~V.} \bibnamefont{Shadrivov}},
  \bibinfo{author}{\bibfnamefont{A.~E.} \bibnamefont{Miroshnichenko}},
  \bibinfo{author}{\bibfnamefont{A.~I.} \bibnamefont{Smirnov}},
  \bibnamefont{and} \bibinfo{author}{\bibfnamefont{Y.~S.}
  \bibnamefont{Kivshar}}, \bibinfo{journal}{Phys. Rev. B}
  \textbf{\bibinfo{volume}{90}}, \bibinfo{pages}{035412}
  (\bibinfo{year}{2014}).

\bibitem[{\citenamefont{Yao et~al.}(2014)\citenamefont{Yao, Tokman, and
  Belyanin}}]{Yao14}
\bibinfo{author}{\bibfnamefont{X.}~\bibnamefont{Yao}},
  \bibinfo{author}{\bibfnamefont{M.}~\bibnamefont{Tokman}}, \bibnamefont{and}
  \bibinfo{author}{\bibfnamefont{A.}~\bibnamefont{Belyanin}},
  \bibinfo{journal}{Phys. Rev. Lett.} \textbf{\bibinfo{volume}{112}},
  \bibinfo{pages}{055501} (\bibinfo{year}{2014}).

\bibitem[{\citenamefont{Cox and {de Abajo}}(2014)}]{CoxAbajo14}
\bibinfo{author}{\bibfnamefont{J.~D.} \bibnamefont{Cox}} \bibnamefont{and}
  \bibinfo{author}{\bibfnamefont{F.~J.~G.} \bibnamefont{{de Abajo}}},
  \bibinfo{journal}{Nat. Commun.} \textbf{\bibinfo{volume}{5}},
  \bibinfo{pages}{5725} (\bibinfo{year}{2014}).

\bibitem[{\citenamefont{Cheng et~al.}(2015)\citenamefont{Cheng, Vermeulen, and
  Sipe}}]{Cheng15}
\bibinfo{author}{\bibfnamefont{J.~L.} \bibnamefont{Cheng}},
  \bibinfo{author}{\bibfnamefont{N.}~\bibnamefont{Vermeulen}},
  \bibnamefont{and} \bibinfo{author}{\bibfnamefont{J.~E.} \bibnamefont{Sipe}},
  \bibinfo{journal}{Phys. Rev. B} \textbf{\bibinfo{volume}{91}},
  \bibinfo{pages}{235320} (\bibinfo{year}{2015}).

\bibitem[{\citenamefont{Cheng et~al.}(2016)\citenamefont{Cheng, Vermeulen, and
  Sipe}}]{Cheng16}
\bibinfo{author}{\bibfnamefont{J.~L.} \bibnamefont{Cheng}},
  \bibinfo{author}{\bibfnamefont{N.}~\bibnamefont{Vermeulen}},
  \bibnamefont{and} \bibinfo{author}{\bibfnamefont{J.~E.} \bibnamefont{Sipe}},
  \bibinfo{journal}{Phys. Rev. B} \textbf{\bibinfo{volume}{93}},
  \bibinfo{pages}{039904(E)} (\bibinfo{year}{2016}).

\bibitem[{\citenamefont{Sabbaghi et~al.}(2015)\citenamefont{Sabbaghi, Lee,
  Stauber, and Kim}}]{Sabbaghi15}
\bibinfo{author}{\bibfnamefont{M.}~\bibnamefont{Sabbaghi}},
  \bibinfo{author}{\bibfnamefont{H.-W.} \bibnamefont{Lee}},
  \bibinfo{author}{\bibfnamefont{T.}~\bibnamefont{Stauber}}, \bibnamefont{and}
  \bibinfo{author}{\bibfnamefont{K.~S.} \bibnamefont{Kim}},
  \bibinfo{journal}{Phys. Rev. B} \textbf{\bibinfo{volume}{92}},
  \bibinfo{pages}{195429} (\bibinfo{year}{2015}).

\bibitem[{\citenamefont{Savostianova and Mikhailov}(2015)}]{Savostianova15}
\bibinfo{author}{\bibfnamefont{N.~A.} \bibnamefont{Savostianova}}
  \bibnamefont{and} \bibinfo{author}{\bibfnamefont{S.~A.}
  \bibnamefont{Mikhailov}}, \bibinfo{journal}{Appl. Phys. Lett.}
  \textbf{\bibinfo{volume}{107}}, \bibinfo{pages}{181104}
  (\bibinfo{year}{2015}).

\bibitem[{\citenamefont{Cox and {de Abajo}}(2015)}]{CoxAbajo15}
\bibinfo{author}{\bibfnamefont{J.~D.} \bibnamefont{Cox}} \bibnamefont{and}
  \bibinfo{author}{\bibfnamefont{F.~J.~G.} \bibnamefont{{de Abajo}}},
  \bibinfo{journal}{ACS Photonics} \textbf{\bibinfo{volume}{2}},
  \bibinfo{pages}{306} (\bibinfo{year}{2015}).

\bibitem[{\citenamefont{Mikhailov}(2016)}]{Mikhailov16a}
\bibinfo{author}{\bibfnamefont{S.~A.} \bibnamefont{Mikhailov}},
  \bibinfo{journal}{Phys. Rev. B} \textbf{\bibinfo{volume}{93}},
  \bibinfo{pages}{085403} (\bibinfo{year}{2016}).

\bibitem[{\citenamefont{Cox et~al.}(2016)\citenamefont{Cox, Silviero, and {de
  Abajo}}}]{CoxAbajo16}
\bibinfo{author}{\bibfnamefont{J.~D.} \bibnamefont{Cox}},
  \bibinfo{author}{\bibfnamefont{I.}~\bibnamefont{Silviero}}, \bibnamefont{and}
  \bibinfo{author}{\bibfnamefont{F.~J.~G.} \bibnamefont{{de Abajo}}},
  \bibinfo{journal}{ACS NANO} \textbf{\bibinfo{volume}{10}},
  \bibinfo{pages}{1995} (\bibinfo{year}{2016}).

\bibitem[{\citenamefont{Marini et~al.}(2016)\citenamefont{Marini, Cox, and {de
  Abajo}}}]{MariniAbajo16}
\bibinfo{author}{\bibfnamefont{A.}~\bibnamefont{Marini}},
  \bibinfo{author}{\bibfnamefont{J.~D.} \bibnamefont{Cox}}, \bibnamefont{and}
  \bibinfo{author}{\bibfnamefont{F.~J.~G.} \bibnamefont{{de Abajo}}},
  \bibinfo{journal}{arXiv:1605.06499}  (\bibinfo{year}{2016}).

\bibitem[{\citenamefont{Mikhailov}(2013)}]{Mikhailov13a}
\bibinfo{author}{\bibfnamefont{S.~A.} \bibnamefont{Mikhailov}},
  \bibinfo{journal}{Phys. Rev. B} \textbf{\bibinfo{volume}{87}},
  \bibinfo{pages}{115405} (\bibinfo{year}{2013}).

\bibitem[{\citenamefont{Moskalenko and Mikhailov}(2014)}]{Moskalenko14}
\bibinfo{author}{\bibfnamefont{A.~S.} \bibnamefont{Moskalenko}}
  \bibnamefont{and} \bibinfo{author}{\bibfnamefont{S.~A.}
  \bibnamefont{Mikhailov}}, \bibinfo{journal}{J. Appl. Phys.}
  \textbf{\bibinfo{volume}{115}}, \bibinfo{pages}{203110}
  (\bibinfo{year}{2014}).

\bibitem[{\citenamefont{Bandhu et~al.}(2013)\citenamefont{Bandhu, Lawton, and
  Nash}}]{Bandhu13}
\bibinfo{author}{\bibfnamefont{L.}~\bibnamefont{Bandhu}},
  \bibinfo{author}{\bibfnamefont{L.~M.} \bibnamefont{Lawton}},
  \bibnamefont{and} \bibinfo{author}{\bibfnamefont{G.~R.} \bibnamefont{Nash}},
  \bibinfo{journal}{Appl. Phys. Lett.} \textbf{\bibinfo{volume}{103}},
  \bibinfo{pages}{133101} (\bibinfo{year}{2013}).

\bibitem[{\citenamefont{Bandhu and Nash}(2014)}]{Bandhu14}
\bibinfo{author}{\bibfnamefont{L.}~\bibnamefont{Bandhu}} \bibnamefont{and}
  \bibinfo{author}{\bibfnamefont{G.~R.} \bibnamefont{Nash}},
  \bibinfo{journal}{Appl. Phys. Lett.} \textbf{\bibinfo{volume}{105}},
  \bibinfo{pages}{263106} (\bibinfo{year}{2014}).

\bibitem[{\citenamefont{Preciado et~al.}(2015)\citenamefont{Preciado, Schulein,
  Nguyen, Barroso, Isarraraz, {von Son}, Lu, Michailow, Moller, Klee
  et~al.}}]{Preciado15}
\bibinfo{author}{\bibfnamefont{E.}~\bibnamefont{Preciado}},
  \bibinfo{author}{\bibfnamefont{F.~J.~R.} \bibnamefont{Schulein}},
  \bibinfo{author}{\bibfnamefont{A.~E.} \bibnamefont{Nguyen}},
  \bibinfo{author}{\bibfnamefont{D.}~\bibnamefont{Barroso}},
  \bibinfo{author}{\bibfnamefont{M.}~\bibnamefont{Isarraraz}},
  \bibinfo{author}{\bibfnamefont{G.}~\bibnamefont{{von Son}}},
  \bibinfo{author}{\bibfnamefont{I.-H.} \bibnamefont{Lu}},
  \bibinfo{author}{\bibfnamefont{W.}~\bibnamefont{Michailow}},
  \bibinfo{author}{\bibfnamefont{B.}~\bibnamefont{Moller}},
  \bibinfo{author}{\bibfnamefont{V.}~\bibnamefont{Klee}}, \bibnamefont{et~al.},
  \bibinfo{journal}{Nat. Commun.} \textbf{\bibinfo{volume}{6}},
  \bibinfo{pages}{8593} (\bibinfo{year}{2015}).

\bibitem[{\citenamefont{Sablikov and Chenskii}(1996)}]{Sablikov96}
\bibinfo{author}{\bibfnamefont{V.~A.} \bibnamefont{Sablikov}} \bibnamefont{and}
  \bibinfo{author}{\bibfnamefont{E.~V.} \bibnamefont{Chenskii}},
  \bibinfo{journal}{JETP} \textbf{\bibinfo{volume}{82}}, \bibinfo{pages}{900}
  (\bibinfo{year}{1996}).

\bibitem[{\citenamefont{Kim et~al.}(2004)\citenamefont{Kim, Korotyeyev,
  Kochelap, Klimov, and Woolard}}]{Kim04}
\bibinfo{author}{\bibfnamefont{K.~W.} \bibnamefont{Kim}},
  \bibinfo{author}{\bibfnamefont{V.~V.} \bibnamefont{Korotyeyev}},
  \bibinfo{author}{\bibfnamefont{V.~A.} \bibnamefont{Kochelap}},
  \bibinfo{author}{\bibfnamefont{A.~A.} \bibnamefont{Klimov}},
  \bibnamefont{and} \bibinfo{author}{\bibfnamefont{D.~L.}
  \bibnamefont{Woolard}}, \bibinfo{journal}{J. Appl. Phys.}
  \textbf{\bibinfo{volume}{96}}, \bibinfo{pages}{6488} (\bibinfo{year}{2004}).

\bibitem[{\citenamefont{Castro~Neto et~al.}(2009)\citenamefont{Castro~Neto,
  Guinea, Peres, Novoselov, and Geim}}]{Neto09}
\bibinfo{author}{\bibfnamefont{A.~H.} \bibnamefont{Castro~Neto}},
  \bibinfo{author}{\bibfnamefont{F.}~\bibnamefont{Guinea}},
  \bibinfo{author}{\bibfnamefont{N.~M.~R.} \bibnamefont{Peres}},
  \bibinfo{author}{\bibfnamefont{K.~S.} \bibnamefont{Novoselov}},
  \bibnamefont{and} \bibinfo{author}{\bibfnamefont{A.~K.} \bibnamefont{Geim}},
  \bibinfo{journal}{Rev. Mod. Phys.} \textbf{\bibinfo{volume}{81}},
  \bibinfo{pages}{109} (\bibinfo{year}{2009}).

\bibitem[{\citenamefont{Yang et~al.}(2007)\citenamefont{Yang, Park, Son, Cohen,
  and Louie}}]{Yang07}
\bibinfo{author}{\bibfnamefont{L.}~\bibnamefont{Yang}},
  \bibinfo{author}{\bibfnamefont{C.-H.} \bibnamefont{Park}},
  \bibinfo{author}{\bibfnamefont{Y.-W.} \bibnamefont{Son}},
  \bibinfo{author}{\bibfnamefont{M.~L.} \bibnamefont{Cohen}}, \bibnamefont{and}
  \bibinfo{author}{\bibfnamefont{S.~G.} \bibnamefont{Louie}},
  \bibinfo{journal}{Phys. Rev. Lett.} \textbf{\bibinfo{volume}{99}},
  \bibinfo{pages}{186801} (\bibinfo{year}{2007}).

\bibitem[{\citenamefont{Mikhailov}(1998)}]{Mikhailov98c}
\bibinfo{author}{\bibfnamefont{S.~A.} \bibnamefont{Mikhailov}},
  \bibinfo{journal}{Phys. Rev. B} \textbf{\bibinfo{volume}{58}},
  \bibinfo{pages}{1517} (\bibinfo{year}{1998}).

\bibitem[{\citenamefont{Mermin}(1970)}]{Mermin70}
\bibinfo{author}{\bibfnamefont{N.~D.} \bibnamefont{Mermin}},
  \bibinfo{journal}{Phys. Rev. B} \textbf{\bibinfo{volume}{1}},
  \bibinfo{pages}{2362} (\bibinfo{year}{1970}).

\bibitem[{\citenamefont{Kragler and Thomas}(1980)}]{Kragler80}
\bibinfo{author}{\bibfnamefont{R.}~\bibnamefont{Kragler}} \bibnamefont{and}
  \bibinfo{author}{\bibfnamefont{H.}~\bibnamefont{Thomas}},
  \bibinfo{journal}{Z. Phys. B} \textbf{\bibinfo{volume}{39}},
  \bibinfo{pages}{99} (\bibinfo{year}{1980}).

\bibitem[{\citenamefont{Stern}(1967)}]{Stern67}
\bibinfo{author}{\bibfnamefont{F.}~\bibnamefont{Stern}},
  \bibinfo{journal}{Phys. Rev. Lett.} \textbf{\bibinfo{volume}{18}},
  \bibinfo{pages}{546} (\bibinfo{year}{1967}).

\bibitem[{\citenamefont{Barreiro et~al.}(2009)\citenamefont{Barreiro, Lazzeri,
  Moser, Mauri, and Bachtold}}]{Barreiro09}
\bibinfo{author}{\bibfnamefont{A.}~\bibnamefont{Barreiro}},
  \bibinfo{author}{\bibfnamefont{M.}~\bibnamefont{Lazzeri}},
  \bibinfo{author}{\bibfnamefont{J.}~\bibnamefont{Moser}},
  \bibinfo{author}{\bibfnamefont{F.}~\bibnamefont{Mauri}}, \bibnamefont{and}
  \bibinfo{author}{\bibfnamefont{A.}~\bibnamefont{Bachtold}},
  \bibinfo{journal}{Phys. Rev. Lett.} \textbf{\bibinfo{volume}{103}},
  \bibinfo{pages}{076601} (\bibinfo{year}{2009}).

\bibitem[{\citenamefont{Mayorov et~al.}(2011)\citenamefont{Mayorov, Gorbachev,
  Morozov, Britnell, Jalil, Ponomarenko, Blake, Novoselov, Watanabe, Taniguchi
  et~al.}}]{Mayorov11b}
\bibinfo{author}{\bibfnamefont{A.~S.} \bibnamefont{Mayorov}},
  \bibinfo{author}{\bibfnamefont{R.~V.} \bibnamefont{Gorbachev}},
  \bibinfo{author}{\bibfnamefont{S.~V.} \bibnamefont{Morozov}},
  \bibinfo{author}{\bibfnamefont{L.}~\bibnamefont{Britnell}},
  \bibinfo{author}{\bibfnamefont{R.}~\bibnamefont{Jalil}},
  \bibinfo{author}{\bibfnamefont{L.~A.} \bibnamefont{Ponomarenko}},
  \bibinfo{author}{\bibfnamefont{P.}~\bibnamefont{Blake}},
  \bibinfo{author}{\bibfnamefont{K.~S.} \bibnamefont{Novoselov}},
  \bibinfo{author}{\bibfnamefont{K.}~\bibnamefont{Watanabe}},
  \bibinfo{author}{\bibfnamefont{T.}~\bibnamefont{Taniguchi}},
  \bibnamefont{et~al.}, \bibinfo{journal}{Nano Lett.}
  \textbf{\bibinfo{volume}{11}}, \bibinfo{pages}{2398} (\bibinfo{year}{2011}).

\bibitem[{\citenamefont{Wixforth et~al.}(1986)\citenamefont{Wixforth, Kotthaus,
  and Weimann}}]{Wixforth86}
\bibinfo{author}{\bibfnamefont{A.}~\bibnamefont{Wixforth}},
  \bibinfo{author}{\bibfnamefont{J.~P.} \bibnamefont{Kotthaus}},
  \bibnamefont{and} \bibinfo{author}{\bibfnamefont{G.}~\bibnamefont{Weimann}},
  \bibinfo{journal}{Phys. Rev. Lett.} \textbf{\bibinfo{volume}{56}},
  \bibinfo{pages}{2104} (\bibinfo{year}{1986}).

\bibitem[{\citenamefont{Falko et~al.}(1993)\citenamefont{Falko, Meshkov, and
  Iordanskii}}]{Falko93}
\bibinfo{author}{\bibfnamefont{V.~I.} \bibnamefont{Falko}},
  \bibinfo{author}{\bibfnamefont{S.~V.} \bibnamefont{Meshkov}},
  \bibnamefont{and} \bibinfo{author}{\bibfnamefont{S.~V.}
  \bibnamefont{Iordanskii}}, \bibinfo{journal}{Phys. Rev. B}
  \textbf{\bibinfo{volume}{47}}, \bibinfo{pages}{9910} (\bibinfo{year}{1993}).

\bibitem[{\citenamefont{Smith and Purcell}(1953)}]{Smith53}
\bibinfo{author}{\bibfnamefont{S.~J.} \bibnamefont{Smith}} \bibnamefont{and}
  \bibinfo{author}{\bibfnamefont{E.~M.} \bibnamefont{Purcell}},
  \bibinfo{journal}{Phys. Rev.} \textbf{\bibinfo{volume}{92}},
  \bibinfo{pages}{1069} (\bibinfo{year}{1953}).

\bibitem[{\citenamefont{Tsui et~al.}(1980)\citenamefont{Tsui, Gornik, and
  Logan}}]{Tsui80}
\bibinfo{author}{\bibfnamefont{D.~C.} \bibnamefont{Tsui}},
  \bibinfo{author}{\bibfnamefont{E.}~\bibnamefont{Gornik}}, \bibnamefont{and}
  \bibinfo{author}{\bibfnamefont{R.~A.} \bibnamefont{Logan}},
  \bibinfo{journal}{Solid State Commun.} \textbf{\bibinfo{volume}{35}},
  \bibinfo{pages}{875} (\bibinfo{year}{1980}).

\bibitem[{\citenamefont{Hirakawa et~al.}(1995)\citenamefont{Hirakawa, Yamanaka,
  Grayson, and Tsui}}]{Hirakawa95}
\bibinfo{author}{\bibfnamefont{K.}~\bibnamefont{Hirakawa}},
  \bibinfo{author}{\bibfnamefont{K.}~\bibnamefont{Yamanaka}},
  \bibinfo{author}{\bibfnamefont{M.}~\bibnamefont{Grayson}}, \bibnamefont{and}
  \bibinfo{author}{\bibfnamefont{D.~C.} \bibnamefont{Tsui}},
  \bibinfo{journal}{Appl. Phys. Lett.} \textbf{\bibinfo{volume}{67}},
  \bibinfo{pages}{2326} (\bibinfo{year}{1995}).

\bibitem[{\citenamefont{Otsuji et~al.}(2011)\citenamefont{Otsuji, Watanabe, E.,
  H., Komori, Satou, Suemitsu, Suemitsu, Sano, Knap et~al.}}]{Otsuji11}
\bibinfo{author}{\bibfnamefont{T.}~\bibnamefont{Otsuji}},
  \bibinfo{author}{\bibfnamefont{T.}~\bibnamefont{Watanabe}},
  \bibinfo{author}{\bibfnamefont{M.~A.} \bibnamefont{E.}},
  \bibinfo{author}{\bibfnamefont{K.}~\bibnamefont{H.}},
  \bibinfo{author}{\bibfnamefont{T.}~\bibnamefont{Komori}},
  \bibinfo{author}{\bibfnamefont{A.}~\bibnamefont{Satou}},
  \bibinfo{author}{\bibfnamefont{T.}~\bibnamefont{Suemitsu}},
  \bibinfo{author}{\bibfnamefont{M.}~\bibnamefont{Suemitsu}},
  \bibinfo{author}{\bibfnamefont{E.}~\bibnamefont{Sano}},
  \bibinfo{author}{\bibfnamefont{W.}~\bibnamefont{Knap}}, \bibnamefont{et~al.},
  \bibinfo{journal}{J. Infrared Milli. Terahz. Waves}
  \textbf{\bibinfo{volume}{32}}, \bibinfo{pages}{629} (\bibinfo{year}{2011}).

\bibitem[{\citenamefont{Otsuji et~al.}(2012)\citenamefont{Otsuji, Tombet,
  Satou, Fukidome, Suemitsu, Sano, Popov, Ryzhii, and Ryzhii}}]{Otsuji12}
\bibinfo{author}{\bibfnamefont{T.}~\bibnamefont{Otsuji}},
  \bibinfo{author}{\bibfnamefont{S.~A.~B.} \bibnamefont{Tombet}},
  \bibinfo{author}{\bibfnamefont{A.}~\bibnamefont{Satou}},
  \bibinfo{author}{\bibfnamefont{H.}~\bibnamefont{Fukidome}},
  \bibinfo{author}{\bibfnamefont{M.}~\bibnamefont{Suemitsu}},
  \bibinfo{author}{\bibfnamefont{E.}~\bibnamefont{Sano}},
  \bibinfo{author}{\bibfnamefont{V.}~\bibnamefont{Popov}},
  \bibinfo{author}{\bibfnamefont{M.}~\bibnamefont{Ryzhii}}, \bibnamefont{and}
  \bibinfo{author}{\bibfnamefont{V.}~\bibnamefont{Ryzhii}},
  \bibinfo{journal}{J. Phys. D: Appl. Phys.} \textbf{\bibinfo{volume}{45}},
  \bibinfo{pages}{303001} (\bibinfo{year}{2012}).

\end{thebibliography}

\end{document}